\documentclass[12pt]{article}
\usepackage{fancyhdr}

\usepackage{mathrsfs}
\usepackage[T1]{fontenc}
\usepackage{mathpazo}
\usepackage{setspace}
\usepackage{amsfonts}
\usepackage{amssymb}
\usepackage{amsmath}
\usepackage{epsfig}
\usepackage{latexsym}
\usepackage{color}
\usepackage{graphicx}
\usepackage{nicefrac}
\usepackage[latin1]{inputenc}
\usepackage{slashed}
\usepackage{multirow}
\usepackage{comment}
\usepackage{soul}
\usepackage{hyperref}
\usepackage[nosort]{cite}
\usepackage{datetime}

\usepackage{braket}

\usepackage{tikz}
\usepackage{capt-of}
\usetikzlibrary{decorations.markings}

\usepackage[titletoc]{appendix}
\usepackage[normalem]{ulem}

\def\hybrid{\topmargin -20pt    \oddsidemargin 0pt
        \headheight 0pt \headsep 0pt
        \textwidth 6.25in       
        \textheight 9.25in       
        \marginparwidth .875in
        \parskip 5pt plus 1pt   \jot = 1.5ex}

\hybrid

\def\baselinestretch{1.2}

\catcode`\@=11

\def\marginnote#1{}
\newcount\hour
\newcount\minute
\newtoks\amorpm
\hour=\time\divide\hour by60
\minute=\time{\multiply\hour by60 \global\advance\minute by-\hour}
\edef\standardtime{{\ifnum\hour<12 \global\amorpm={am}%
        \else\global\amorpm={pm}\advance\hour by-12 \fi
        \ifnum\hour=0 \hour=12 \fi
        \number\hour:\ifnum\minute<10 0\fi\number\minute\the\amorpm}}
\edef\militarytime{\number\hour:\ifnum\minute<10 0\fi\number\minute}

\def\draftlabel#1{{\@bsphack\if@filesw {\let\thepage\relax
   \xdef\@gtempa{\write\@auxout{\string
      \newlabel{#1}{{\@currentlabel}{\thepage}}}}}\@gtempa
   \if@nobreak \ifvmode\nobreak\fi\fi\fi\@esphack}
        \gdef\@eqnlabel{#1}}
\def\@eqnlabel{}
\def\@vacuum{}
\def\draftmarginnote#1{\marginpar{\raggedright\scriptsize\tt#1}}

\def\draft{\oddsidemargin -.5truein
        \def\@oddfoot{\sl preliminary draft \hfil
        \rm\thepage\hfil\sl\today\quad\militarytime}
        \let\@evenfoot\@oddfoot \overfullrule 3pt
        \let\label=\draftlabel
        \let\marginnote=\draftmarginnote
   \def\@eqnnum{(\theequation)\rlap{\kern\marginparsep\tt\@eqnlabel}%
\global\let\@eqnlabel\@vacuum}  }

\def\preprint{\twocolumn\sloppy\flushbottom\parindent 2em
        \leftmargini 2em\leftmarginv .5em\leftmarginvi .5em
        \oddsidemargin -.5in    \evensidemargin -.5in
        \columnsep .4in \footheight 0pt
        \textwidth 10.in        \topmargin  -.4in
        \headheight 12pt \topskip .4in
        \textheight 6.9in \footskip 0pt
        \def\@oddhead{\thepage\hfil\addtocounter{page}{1}\thepage}
        \let\@evenhead\@oddhead \def\@oddfoot{} \def\@evenfoot{} }

\def\numberbysection{\@addtoreset{equation}{section}
        \def\theequation{\thesection.\arabic{equation}}}

\def\underline#1{\relax\ifmmode\@@underline#1\else
        $\@@underline{\hbox{#1}}$\relax\fi}

\def\titlepage{\@restonecolfalse\if@twocolumn\@restonecoltrue\onecolumn
     \else \newpage \fi \thispagestyle{empty}\c@page\z@
        \def\thefootnote{\fnsymbol{footnote}} }

\def\endtitlepage{\if@restonecol\twocolumn \else \newpage \fi
        \def\thefootnote{\arabic{footnote}}
        \setcounter{footnote}{0}}  

\catcode`@=12
\relax

\def\figcap{\section*{Figure Captions\markboth
        {FIGURECAPTIONS}{FIGURECAPTIONS}}\list
        {Figure \arabic{enumi}:\hfill}{\settowidth\labelwidth{Figure
999:}
        \leftmargin\labelwidth
        \advance\leftmargin\labelsep\usecounter{enumi}}}
 \relax
\def\tablecap{\section*{Table Captions\markboth
        {TABLECAPTIONS}{TABLECAPTIONS}}\list
        {Table \arabic{enumi}:\hfill}{\settowidth\labelwidth{Table
999:}
        \leftmargin\labelwidth
        \advance\leftmargin\labelsep\usecounter{enumi}}}
 \relax
\def\reflist{\section*{References\markboth
        {REFLIST}{REFLIST}}\list
        {[\arabic{enumi}]\hfill}{\settowidth\labelwidth{[999]}
        \leftmargin\labelwidth
        \advance\leftmargin\labelsep\usecounter{enumi}}}
 \relax
\makeatletter
\newcounter{pubctr}
\def\publist{\@ifnextchar[{\@publist}{\@@publist}}
\def\@publist[#1]{\list
        {[\arabic{pubctr}]\hfill}{\settowidth\labelwidth{[999]}
        \leftmargin\labelwidth
        \advance\leftmargin\labelsep
        \@nmbrlisttrue\def\@listctr{pubctr}
        \setcounter{pubctr}{#1}\addtocounter{pubctr}{-1}}}
\def\@@publist{\list
        {[\arabic{pubctr}]\hfill}{\settowidth\labelwidth{[999]}
        \leftmargin\labelwidth
        \advance\leftmargin\labelsep
        \@nmbrlisttrue\def\@listctr{pubctr}}}
 \relax
\makeatother
\newskip\humongous \humongous=0pt plus 1000pt minus 1000pt

\newif\ifdtup

\relax

\def\be{\begin{equation}}
\def\ee{\end{equation}}
\def\ba{\begin{eqnarray}}
\def\ea{\end{eqnarray}}

\def\del{\partial}

\def\k{\kappa}
\def\r{\rho}
\def\a{\alpha}

\def\b{\beta}

\def\g{\gamma}

\def\d{\delta}
\def\D{\Delta}
\def\e{\epsilon}

\def\th{\theta}

\def\m{\mu}
\def\n{\nu}

\def\l{\lambda}

\def\s{\sigma}

\def\cN{{\cal N}}

\def\cM{{\cal M}}

\def\cM{{\cal M}}
\def\cN{{\cal N}}

\def\cS{{\cal S}}

\def\no{\noindent}

\def\qq{\qquad}

\def\IR{\relax{\rm I\kern-.18em R}}

\def \ha {{1\over 2}}

\def \ov {\over}

\def\IR{\relax{\rm I\kern-.18em R}}
\def\IL{\relax{\rm I\kern-.18em L}}

\def\inv{^{\raise.15ex\hbox{${\scriptscriptstyle -}$}\kern-.05em 1}}

\def\cM{{\cal M}}

\begin{document}

\newcommand{\eqn}[1]{(\ref{#1})}

\renewcommand{\theequation}{\thesection.\arabic{equation}}
\csname @addtoreset\endcsname{equation}{section}

\begin{titlepage}
\begin{center}

\phantom{xx}
\vskip 0.5in


{\large \bf On the stability of $AdS$ backgrounds with $\l$-deformed factors}

\vskip 0.5in

{\bf G. Itsios}${}^{1a}$, \hskip .15cm 
{\bf P. Panopoulos}${}^{1b}$, \hskip .15cm 
{\bf K. Sfetsos}${}^{1c}$\hskip .13cm and \hskip .12cm 
{\bf D. Zoakos}${}^{1,2d}$ 
\vskip 0.1in

{\em 
${}^1$ Department of Nuclear and Particle Physics, \\
Faculty of Physics, 
National and Kapodistrian University of Athens, \\
Athens 15784, Greece. \\

\vskip .2 cm 
${}^2$ Department of Engineering and Informatics,\\
 Hellenic American University, \\
436 Amherst st, Nashua, NH 03063 USA. \\
}

\vskip .2in


\end{center}

\vskip .4in

\centerline{\bf Abstract}

\noindent
We investigate the stability of the non-supersymmetric solutions of type-IIB supergravity
having an unwarped $AdS$ factor and $\lambda$-deformed subspaces found in \cite{Itsios:2019izt}.
Among the plethora of solutions we study the perturbative stability of backgrounds with an $AdS_n$, with $n=3,4,6$, factor. Our analysis is performed  from a lower dimensional effective theory which we construct.
We uncover the regions and isolated points in the parameter space of potential perturbative stability.

\no

\vfill
\no
 {\footnotesize
$^a$  gitsios@phys.uoa.gr, \\
$^b$ ppanopoulos@phys.uoa.gr, \\
$^c$ ksfetsos@phys.uoa.gr, \\
$^d$ zoakos@gmail.com \&  dzoakos@phys.uoa.gr 
}

\end{titlepage}
\vfill
\eject

\newpage

\tableofcontents

\noindent

\def\baselinestretch{1.2}
\baselineskip 20 pt
\noindent


\setcounter{equation}{0}

\section{Introduction}

In recent years novel two-dimensional quantum field theories with a group theoretical basis and remarkable properties have been constructed. One such class comprises theories that generically come under the name of $\l$-deformations 
\cite{Sfetsos:2013wia, Georgiou:2016zyo,Georgiou:2017jfi,Georgiou:2018hpd,Georgiou:2018gpe,Driezen:2019ykp,Hollowood:2014rla,Hollowood:2014qma,Sfetsos:2017sep}. They represent finite, integrable in specific cases, deformations of WZW or gauged WZW theories.
In parallel with the above development, another related class of integrable $\s$-models \cite{Klimcik:2002zj,Klimcik:2008eq,Klimcik:2014, Delduc:2013fga,Delduc:2013qra,Arutyunov:2013ega}, ultimately connected to the Principal Chiral Model (PCM) for groups and cosets and generically known as  $\eta$-deformations has also been constructed. 

While these works have been further developed in various directions,
in the present work we are focused in aspects concerning their embedding to type-II supergravity. 
In general this is a challenging task to perform and necessarily 
involves turning on a dilaton as well as Ramond-Ramond (RR) fields that supplement the two-dimensional $\s$-model  metric and antisymmetric tensor fields.
Most of the examples constructed in the literature 
\cite{Sfetsos:2014cea,Demulder:2015lva,Hoare:2015gda,Borsato:2016zcf,Chervonyi:2016ajp,Borsato:2016ose,Hoare:2015wia,Lunin:2014tsa,Hoare:2018ngg,Seibold:2019dvf}, 
have the entire space-time deformed and therefore they lack an $AdS$ space-time as part of the full supergravity solution.
The only exception is the work of \cite{Itsios:2019izt}, 
where (unwarped) $AdS$ solutions with $\l$-deformed internal subspaces have been explicitly constructed.
 
Applying the machinery of AdS/CFT correspondence is the natural next step in order to enlighten the field theory 
side of the solutions of \cite{Itsios:2019izt}.
However, the two important characteristics of these backgrounds, namely the absence of supersymmetry and the 
presence of an unwarped $AdS$ factor, combined with the Ooguri-Vafa conjecture \cite{Ooguri:2016pdq}, 
suggest that an investigation of their stability, at least perturbatively, is imminent. 
Specifically, this conjecture states that any non-supersymmetric $AdS$ vacuum is unstable and it is a stronger version of the weak gravity conjecture \cite{ArkaniHamed:2006dz}. 
Therefore the tedious effort of checking the stability of the $AdS$ backgrounds  with $\l$-deformed factors 
can be put in the wider context of satisfying or disproving this conjecture. 
What makes the models of \cite{Itsios:2019izt} attractive in this context is their relative simplicity and the fact that the various factors 
correspond to integrable subspaces from a two-dimensional $\s$-model point of view. 
The latter property empirically and intuitively, if not guaranteeing stability, it makes instability less likely to be the case. 
In that respect, we note that the parametric space in some solutions is two-dimensional, leaving enough room for stability in part(s) of it.
However notice that, even if we could prove perturbative stability of the $\l$-deformed backgrounds, 
to disprove the conjecture we should also exclude possible non-perturbative instabilities.

The state of the art in the field of investigating perturbative stability is coming from a method that is based on 
exceptional field theory \cite{Hohm:2013pua} and was developed in
\cite{Malek:2019eaz, Malek:2020yue} for computing Kaluza--Klein
spectra of maximal gauged supergravity vacua.  In \cite{Guarino:2020flh} this method was applied to prove 
perturbative stability for the non-supersymmetric $G_2$-invariant $AdS_4\times S^6$ 
background of massive-IIA supergravity
\footnote{
Another noteworthy non-supersymmetric and seemingly perturbatively stable string theory paradigm
can be found in \cite{Basile:2018irz}. In  \cite{Apruzzi:2019ecr},
$AdS_7$ non-supersymmetric solutions of massive IIA were analyzed and both
perturbative and non-perturbative instabilities were found.}, 
through the analytic computation of the full Kaluza--Klein spectrum.
The aforementioned calculation combined with the absence of a 
brane-jet instability \cite{Bena:2020xxb}, i.e. a recently introduced non-perturbative instability, for these backgrounds 
\cite{Guarino:2020jwv}, sets an interesting challenge for the Ooguri--Vafa conjecture. 

The plan of the paper is as follows: In section \ref{GravityScalars} we present in a minimalistic way the fluctuation 
analysis for a wide class of effective theories containing only gravity and scalars. Since this structure is a common characteristic 
of all the dimensionally reduced solutions that we study in the current paper, we present the results in a 
unified framework. We perform a perturbative analysis around an $AdS$ solution having constant scalar fields.
Analysing stability boils down to determining the eigenvalues of a mass matrix, which is constructed using the 
scalar potential and the constant metric in the scalar field space. 

In section \ref{AdS3solution} we perform a stability analysis of the type-IIB $AdS_3 \times S^3 \times 
CS^2_\l \times CH_{2,\l}$ solution, where $CS^2_\l$ and $CH_{2,\l}$ are the $\lambda$-deformed (two dimensional) 
coset CFTs $SU(2)/U(1)$ and $SL(2,\mathbb{R})/SO(1,1)$, respectively \cite{Sfetsos:2013wia,Itsios:2019izt}. 
We adopt an initial reduction ansatz containing only gravity and scalars which is also consistent with the Bianchi identities and flux equations for the RR-fields. The resulting equations of motions are eventually derived from a lower dimensional effective action of the type 
studied in section \ref{GravityScalars}. From this action we construct the mass matrix.
The analysis  is depicted in figures \ref{BFb2b3} and  \ref{BFb4} and reveals regions of 
stability/instability separated by a critical line between the two. 

In section \ref{AdS4solution} we perform a  stability analysis of the type-IIB 
$AdS_4 \times \cN_4 \times CS^2_\l$ where the four-dimensional space $\cN_4$ can be any of the following spaces 
$(S^4 , H_4 , T^4)$ \cite{Itsios:2019izt}. 
We follow the same line of reasoning as before and arrive to a mass matrix.
In figure \ref{BFAdS4} we plot its eigenvalues as a function of the deformation parameter 
when $\cN_4$ is either $S^4$ or $H_4$. Both cases are proven unstable. 
The case of  ${\cN_4} = T^4$ provides a range of potential stability.

In section \ref{AdS6solution} we perform the stability analysis of the type-IIB 
$AdS_6 \times \cN_2 \times CS^2_\l$ where the two-dimensional space $\cN_2$ can be any of the following
$(S^2 , H_2 , T^2)$ \cite{Itsios:2019izt}. 
In figure \ref{BFAdS6}, we plot the eigenvalues as a function 
of the deformation parameter when $\cN_2$ is either $S^2$ or $H_2$ which reveals instability. 
The case of $T^2$,  existing for zero deformation parameter, is also unstable. 

In section \ref{conclusions} we gather our results and discuss potential future directions. 
The main text is supplemented with two useful appendices. In appendix \ref{IIB}, to set up the basis for the notation, we list the 
equations of motion  of type-IIB supergravity in the string frame. In appendix  \ref{lambda-solutions} we summarise all the 
supergravity solutions of  \cite{Itsios:2019izt} that we analyse in the main text.


\section{Gravity and scalars}
\label{GravityScalars}
\def\gn{\mathfrak{g}}
\def\Rn{\mathfrak{R}}

Our stability analysis of the type-II solutions  will be based on particular dimensional reductions 
of the ten-dimensional fields. These reductions will give rise to lower dimensional theories of gravity coupled to scalars. 
In all cases the background values for the scalars are constants while the background geometry is an $AdS$ space. 
In this section we study the linearized equations of motion up to first order in the fluctuations for this class of effective theories.

Consider a theory of gravity coupled to $n$ scalars in $D$ dimensions described by the action
\begin{equation}
 \label{ActionDdimensions}
 S(\gn, X) = \frac{1}{2 \k^2_D} \int d^D x \sqrt{|\gn|} \Big(  \Rn - \g_{ij} \partial X^i \cdot \partial X^j - V(X)  \Big) \, ,
\end{equation}
where the vector $X = (X_1 , \ldots , X_n)$ encodes the $n$ scalars and $\g_{ij}$ is an $n \times n$ constant symmetric matrix, 
 playing the r\^ole of the metric in the scalar field space.

\no
The Einstein equations arising from varying this action with respect to the metric $\gn_{\m\n}$ are
\begin{equation}
 \Rn_{\m\n} - \g_{ij} \partial_\m X^i \partial_\n X^j - \frac{1}{2} \gn_{\m\n} \Big(  \Rn - \g_{ij} \partial X^i \cdot \partial X^j - V(X) \Big) = 0 \, .
\end{equation}
The trace of the Einstein equations gives
\begin{equation}
 \Rn - \g_{ij} \partial X^i \cdot \partial X^j - \frac{D}{D-2} V(X) = 0 \, .
\end{equation}
Eliminating the Ricci scalar from the Einstein equations we end up with
\begin{equation}
 \label{EinsteinEqsEinsteinFrame}
 \Rn_{\m\n} - \g_{ij} \partial_\m X^i \partial_\n X^j - \frac{\gn_{\m\n}}{D - 2} V(X) = 0 \, .
\end{equation}

\no
Moreover the equations of motion for the scalars are
\begin{equation}
 \label{ScalarEqsEinsteinFrame}
 \nabla^2_\gn X^i - \frac{1}{2} \g^{ij} \partial_j V(X) = 0 \, ,
\end{equation}
where $\g^{ij}$ is the inverse of $\g_{ij}$.

\subsection{Perturbations around $AdS_D$}

We will linearise the equations of motion \eqref{EinsteinEqsEinsteinFrame} and \eqref{ScalarEqsEinsteinFrame} assuming an $AdS_D$ background solution
\begin{equation} 
\label{SolutionGravityScalars}
ds^2_D = \bar{\gn}_{\m\n} dx^\m dx^\n = L^2 \bigg(r^2 \eta_{\a\b} dx^\a dx^\b + \frac{dr^2}{r^2} \bigg)\, , \quad
 \bar{X}^i = \textrm{const} \, ,
\end{equation}
where early Greek indices run through $\a,\b=0,1,\dots,  D-2$ and the Minkowski metric is mostly plus. 
Note the bar above the background metric and scalars, a notation that we will follow in our paper. 

\no
Then, using  \eqref{SolutionGravityScalars} in \eqref{EinsteinEqsEinsteinFrame} and \eqref{ScalarEqsEinsteinFrame} implies that
\begin{equation}
 \begin{aligned}
  & \bar{\Rn}_{\m\n} = \frac{\bar{\gn}_{\m\n}}{D - 2} V(\bar{X}) = - \frac{D - 1}{L^2} \bar{\gn}_{\m\n} \quad \Longrightarrow \quad V(\bar{X}) = - \frac{(D - 1) (D - 2)}{L^2} \, ,
  \\
  & \partial_i V(\bar{X}) = 0 \, , \quad i = 1 , \ldots , n \, .
 \end{aligned}
\end{equation}
We will now consider the fluctuations
\begin{equation}
 \gn_{\m\n} = \bar{\gn}_{\m\n} + \d \gn_{\m\n} \, , \qquad X^i = \bar{X}^i + \d X^i \, , \qquad i = 1 , \ldots , n \, .
\end{equation}
Using the above we next study the perturbations of \eqref{EinsteinEqsEinsteinFrame} and \eqref{ScalarEqsEinsteinFrame} to linear order.

\subsubsection{Metric fluctuations}

The linearized version \eqref{EinsteinEqsEinsteinFrame} around \eqn{SolutionGravityScalars} reads
\begin{equation}
 \d \Rn_{\m\n} - \frac{V(\bar{X})}{D - 2} \d \gn_{\m\n} = 0 \quad \Longrightarrow \quad \d \Rn_{\m\n} + \frac{D - 1}{L^2} \d \gn_{\m\n} = 0\, ,
\end{equation}
or more explicitly that
\begin{equation}
 \label{LinearizedEinsteinEqGravityScalars}
  \nabla^\s \nabla_\m \d \gn_{\s\n} 
  + \nabla^\s \nabla_\n \d \gn_{\m\s} 
  - \nabla^2 \d \gn_{\m\n} 
  - \bar{\gn}^{\r\s} \nabla_\n \nabla_\m \d \gn_{\r\s}
  + 2 \, \frac{D - 1}{L^2} \d \gn_{\m\n} = 0 \, .
\end{equation}
We choose the transverse-traceless gauge
\begin{equation}
 \nabla^\m \d \gn_{\m\n} = 0 \, , \qquad \gn^{\m\n} \, \d \gn_{\m\n} = 0 \, ,
\end{equation}
which is possible since our action \eqn{ActionDdimensions} starts  with the standard Einstein--Hilbert term.
From the definition of the Riemann tensor we have that 
\begin{equation}
[\nabla_\r, \nabla_\s] \d \gn_{\m\n} =  - \bar{\Rn}^\l_{\,\,\,\, \m \r \s} \d \gn_{\l\n} - \bar{\Rn}^\l_{\,\,\,\, \n \r \s} \d \gn_{\m\l} \, .
\end{equation}
Therefore, using the transversality condition we have that 
\begin{equation}
 \nabla^\r \nabla_\s \d \gn_{\r\n} = \bar{\Rn}^\l_{\,\,\,\, \s} \d \gn_{\l\n} + \bar{\Rn}_{\l \n \r \s} \d \gn^{\r\l} \, .
\end{equation}
In addition, the Riemann tensor of the $AdS_D$ geometry is
\begin{equation}
 \bar{\Rn}_{\l \n \r \s} = \frac{1}{L^2}\big(  \gn_{\l\s} \gn_{\n\r} - \gn_{\l\r} \gn_{\n\s} \big) \quad \Longrightarrow \quad \bar{\Rn}_{\n\s} = - \frac{D - 1}{L^2} \gn_{\n\s} \, .
\end{equation}
Using the above properties
\begin{equation}
 \nabla^\r \nabla_\s \d \gn_{\r\n} = - \frac{D }{L^2} \d \gn_{\s\n} \, .
\end{equation}
 From this and the transverse-traceless gauge, \eqref{LinearizedEinsteinEqGravityScalars} becomes
\begin{equation}
 \label{EqMasslesGraviton}
  \nabla^2 \d \gn_{\m\n}
  + \frac{2}{L^2} \d \gn_{\m\n} = 0\ ,
\end{equation}
which is the equation for a massless graviton in $AdS_D$. We conclude 
that the metric fluctuations are stable under small perturbations. Any potential instability, will arise solely from the particular form  
of the scalar potential $V(X)$, that enters the analysis of the scalar fluctuations. 


\subsubsection{Scalar fluctuations}

We now move to the perturbations of \eqref{ScalarEqsEinsteinFrame}. To linear order these read
\begin{equation}
 \nabla^2_{\bar{\gn}} \d X^i - \big(  M^2 \big)^i{}_j \d X^j = 0 \, .
\end{equation}
%
%
where we have defined the mass squared matrix
\begin{equation}
 \label{M2matrix}
 \big(M^2\big)^i{}_j = \frac{1}{2} \g^{ik} \partial_j \partial_k V(X)\Big |_{X=\bar X}\ .
\end{equation}
Suppose now that we find a matrix $P$ diagonalising $M^2$, i.e. 
\begin{equation}
 P M^2 P^{-1} = \textrm{diag}(d_1 , \ldots , d_n) \, ,
\end{equation}
where $d_i$ with $i = 1 , \ldots , n$ are the eigenvalues of $M^2$. Note that it is not warranted that these eigenvalues are real 
 since, for instance, $M^2$ is not necessarily a Hermitian matrix. 
Continuing the analysis, the equation we want to solve reduces to
\begin{equation}
 \nabla^2_{\bar{\gn}} f_i - d_i \, f_i = 0 \, , \qquad i = 1 , \ldots , n \, ,
\end{equation}
with $P \d X = (f_1 , \ldots , f_n)$. If we assume for $f_i$ 
a plane wave dependence on the coordinates $x^\a$, i.e. $e^{i k^{(i)}_\a x^\a}$ and expand the Laplacian we find that 
%
%
%
\begin{equation}
 \partial^2_r f_i + \frac{D}{r} \partial_r f_i - \Big(  \frac{L^2 \, d_i}{r^2} - \frac{m^2_i}{r^4}  \Big) f_i = 0 \, , \qquad i = 1 , \ldots , n \, ,
\end{equation}
where $m_i^2 = - k^{(i)}\cdot k^{(i)}$.
We will look the behaviour of $f_i$ with $i = 1 , \ldots , n$ for large $r$. Thus we assume $f_i \sim r^{- \D_i}$. If we plug this into the above differential equation and keep the dominant terms we end up with an algebraic equation for $\D_i$ which is
\begin{equation}
\label{massdim}
 \D_i (\D_i + 1 - D) = L^2 \, d_i \, , \quad  i = 1 , 2\ldots , n \, .
\end{equation}
This is the well known mass-dimension formula obtained in the context of the AdS/CFT correspondence in \cite{Witten:1998qj}.
Reality of the scaling dimensions $\D_i$ requires that
\begin{equation}
 \label{BFbound}
 d_i \geqslant - \Big(  \frac{D - 1}{2 L}  \Big)^2 \, , \quad \forall\ i = 1 , \ldots , n \, ,
\end{equation}
which is known as the \emph{Breitenlohner-Freedman} (BF) bound \cite{Breitenlohner:1982jf}.


\section{The $AdS_3$ solution}
\label{AdS3solution}

In this section we examine  the stability of the type-IIB solution on $AdS_3 \times S^3 \times CS^2_\l \times CH_{2,\l}$, 
where we use the notation $CS^2_\l$ and $CH_{2,\l}$ for the $\lambda$-deformed cosets $SU(2)/U(1)$ and $SL(2,\mathbb{R})/SO(1,1)$, respectively. 
 The solution depends on the deformation parameter $\l$, the scale $\ell$ of the $AdS_3$ and the level $k$ of the undeformed CFTs. 
It turns out that, the parametric space of the mass matrix arising from the stability analysis is two-dimensional since the $k$ and $\ell$ will appear via the combination $\hat \ell = k \ell$.
Details of the solution can be found in appendix \ref{AdS3}. We will show that demanding perturbative stability restricts the allowed 
parametric space. 

\subsection{The reduction ansatz}

We start by introducing a reduction ansatz for the solution in appendix \ref{AdS3}. The reduction will be along the three-sphere and the two $\l$-deformed spaces of the ten-dimensional space \eqref{solAdS3metric}. Thus, for the metric we adopt the 
following ansatz (our approach is similar  in spirit to that in \cite{Liu:2010sa, Gauntlett:2010vu,Skenderis:2010vz})
\ba
 \label{AnsatzAdS3metric}
 &&  d\hat{s}^2  =  e^{2A} \Big[   ds^2_{\cM_3} + e^{2 \psi} \big(  d\th^2_1 + \sin^2\th_1 \, d\th^2_2 + \sin^2\th_1 \sin^2\th_2 \, d\th^2_2 \big)
 \nonumber
 \\
 &&\quad 
  + e^{2 \phi_y} \Big(  \l^2_+ \, e^{2 \chi_1} \, dy^2_1 + \l^2_- \, e^{2 \chi_2} \, dy^2_2  \Big)
  + e^{2 \phi_z} \Big(  \l^2_+ \, e^{2 \chi_3} \, dz^2_1 + \l^2_- \, e^{2 \chi_4} \, dz^2_2  \Big)  \Big] \, ,
\ea
where $\cM_3$ is a three-dimensional space with metric
\begin{equation}
 ds^2_{\cM_3} = g_{\m\n} \, dx^\m \, dx^\n\ .
\end{equation}
The scalars $A , \psi , \chi_1 , \ldots , \chi_4$ can only depend on the coordinates $x^\m$ of $\cM_3$. For the NS three-form and the dilaton we take
\be
 \label{AnsatzAdS3NS}
   \widehat{H}_3 = 0 \, , \qquad \widehat{\Phi} (x,y,z) = 4 A(x) + \phi_y(y) + \phi_z(z) \, ,
\ee
where $\phi_y(y)$ and $\phi_z(z)$ are given by \eqn{CS2l} and \eqn{CH2l}, respectively. 
Finally, for the RR-sector the ansatz is
\be
 \label{AnsatzAdS3RR}
 \begin{split}
  \widehat{F}_1 & = 0 \, , \qquad\widehat{F}_3 = 0 \, ,
  \\
  \widehat{F}_5 & = dz_1 \wedge dy_2 \wedge \Big(  c_1 e^{\chi_2 - \chi_1 + \chi_3 - \chi_4 - 3 \psi} \, \textrm{Vol}(\cM_3) + c_2 \textrm{Vol}(S^3) \Big)
\\
  & - dz_2 \wedge dy_1 \wedge \Big(  c_2 e^{\chi_1 - \chi_2 - \chi_3 + \chi_4 - 3 \psi} \textrm{Vol}(\cM_3)+ c_1 \textrm{Vol}(S^3) \Big) \, ,
 \end{split}
\ee
where $\textrm{Vol}(\cM_3)$ is the volume form on $\cM_3$ and $\textrm{Vol}(S^3)$ is given explicitly by 
\eqref{VolsAdS3S3}. 
The constants $c_1$ and $c_2$ are 
\begin{equation}
 c_1 = 2 k \Big(  \frac{2}{\ell} \Big)^{\frac{3}{2}} \sqrt{\frac{\ell - \m}{2}} \, , \qquad 
 c_2 = 2 k \Big(  \frac{2}{\ell} \Big)^{\frac{3}{2}} \sqrt{\frac{\ell + \m}{2}} \, .
\end{equation}
The solution of appendix \ref{AdS3} is obtained by taking the space $\cM_3$ to be an $AdS_3$ with line element
\begin{equation}
 \label{BackgrValsAdS3metric}
 ds^2_{AdS_3} = \bar{g}_{\m\n} \, dx^\m \, dx^\n = \frac{2}{\ell} \bigg( r^2 \eta_{\a\b} dx^\a dx^\b  + \frac{dr^2}{r^2} \bigg)\ ,
\end{equation}
normalised as $\bar{R}_{\m\n} = - \ell \, \bar{g}_{\m\n}$ and by setting the scalars to
\begin{equation}
 \label{BackgrValsAdS3scalars}
 \bar A = \bar \chi_1 = \bar \chi_2 = \bar \chi_3 = \bar \chi_4 = 0 \, , \quad \bar \psi = \frac{1}{2} \ln \frac{2}{\ell} \, .
\end{equation}

\subsection{The equations of motion}

 To find the equations of motion for the scalars $A , \psi , \chi_1 , \ldots , \chi_4$ and the metric $g_{\m\n}$ on $\cM_3$ we insert the ansatz \eqref{AnsatzAdS3metric}, \eqref{AnsatzAdS3NS} and \eqref{AnsatzAdS3RR} in the equations of motion of the type-IIB supergravity summarised in appendix \ref{IIB}. We first notice that the equations for the form fields \eqref{BianchiIIB} and \eqref{FluxesIIB} are trivially satisfied. Therefore, we are only concerned with the dilaton and Einstein equations \eqref{DilatonEinstein}. 

\no
Tensors constructed below and all contractions are performed with respect to the metric $g_{\m\n}$ on $\cM_3$.

\paragraph{The dilaton equation:}
The dilaton equation \eqref{DilatonEinstein} reduces to
\ba
 \label{DilatonReductionAdS3}
  && R + 6 \, e^{- 2 \psi}
  - 2 \nabla^2_g \big(  3 \psi + \chi_1 + \chi_2 + \chi_3 + \chi_4  \big)
  - \big(  3 \, \partial \psi + \partial \chi_1 + \partial \chi_2 + \partial \chi_3 + \partial \chi_4  \big)^2
  \nonumber
  \\
  && - 3 \big( \partial \psi \big)^2
  -  \big( \partial \chi_1 \big)^2
  - \big( \partial \chi_2 \big)^2
  - \big( \partial \chi_3 \big)^2
  - \big( \partial \chi_4 \big)^2
  - 2 \, \partial \big(  3 \psi + \chi_1 + \chi_2 + \chi_3 + \chi_4 \big) \cdot \partial A
  \nonumber
    \\
  && - 2 \, \nabla^2_g A
  - 8 \, \big(  \partial A  \big)^2
  + 2 \, \frac{e^{- 2 \chi_1}}{\l^2_+}
  + 2 \, \frac{e^{- 2 \chi_2}}{\l^2_-}
  - 2 \, \frac{e^{- 2 \chi_3}}{\l^2_+}
  - 2 \, \frac{e^{- 2 \chi_4}}{\l^2_-} = 0 \, .
\ea

\paragraph{The directions along $\cM_3$:}
Restricting ourselves to the components of the Einstein equations \eqref{DilatonEinstein} along the $\cM_3$ directions we get
\ba
 \label{EinsteinReductionAdS3}
  && R_{\m\n}
   - \nabla_{\m} \nabla_{\n} \big(  3 \psi + \chi_1 + \chi_2 + \chi_3 + \chi_4  \big)
   - 3 \partial_{\m} \psi \, \partial_{\n} \psi
   - \partial_{\m} \chi_1 \, \partial_{\n} \chi_1
   - \partial_{\m} \chi_2 \, \partial_{\n} \chi_2
     \nonumber
     \\
  && - \partial_{\m} \chi_3 \, \partial_{\n} \chi_3
   - \partial_{\m} \chi_4 \, \partial_{\n} \chi_4
   - 8 \, \partial_\m A \partial_\n A
   - g_{\m\n} \, \nabla^2_g A
   - g_{\m\n} \, \partial \big(  3 \psi + \chi_1 + \chi_2 + \chi_3 + \chi_4 \big) \cdot \partial A
  \nonumber
     \\
  &&\qq = - g_{\m\n} \frac{e^{- 6 \psi}}{4 k^2} \Big(  c^2_1 e^{ - 2 \chi_1 - 2 \chi_4} + c^2_2 e^{ - 2 \chi_2 - 2 \chi_3}  \Big) \, .
\ea

\no
Taking the trace of the equation above and using it in order to eliminate the Ricci scalar from \eqref{DilatonReductionAdS3}
we obtain that
\be
\begin{split}
\label{Constr1AdS3}
 \nabla^2_g & \big(  3 \psi + \chi_1 + \chi_2 + \chi_3 + \chi_4 - A \big)
\\ 
&
+ \partial \big(  3 \psi + \chi_1 + \chi_2 + \chi_3 + \chi_4 \big) \cdot \partial \big(  3 \psi + \chi_1 + \chi_2 + \chi_3 + \chi_4 - A \big)
\\
 & - 6 \, e^{- 2 \psi}
   + 3 \, \frac{e^{- 6 \psi}}{4 k^2} \Big(  c^2_1 e^{ - 2 \chi_1 - 2 \chi_4} + c^2_2 e^{ - 2 \chi_2 - 2 \chi_3}  \Big)
 \\
 &
   - 2 \, \frac{e^{- 2 \chi_1}}{\l^2_+}
   - 2 \, \frac{e^{- 2 \chi_2}}{\l^2_-}
   + 2 \, \frac{e^{- 2 \chi_3}}{\l^2_+}
   + 2 \, \frac{e^{- 2 \chi_4}}{\l^2_-} = 0 \, .
  \end{split}
\ee
It is convenient to use this equation instead of the equivalent one in \eqn{DilatonReductionAdS3}.

\paragraph{The directions along $S^3$:}
It turns out that the only non-vanishing components of the Einstein equations along the sphere directions are the diagonal ones. 
As expected, due to symmetry, we get a single equation given by
\begin{equation}
 \begin{aligned}
  \label{Constr2AdS3}
   & \nabla^2_g \big(   A + \psi  \big)
   + \partial \big( A + \psi \big) \cdot \partial \big(  3 \psi + \chi_1 + \chi_2 + \chi_3 + \chi_4  \big)
   - 2 e^{- 2 \psi}
\\
   &\qq  = - \frac{e^{- 6 \psi}}{4 k^2} \Big(  c^2_1 e^{ - 2 \chi_1 - 2 \chi_4} + c^2_2 e^{ - 2 \chi_2 - 2 \chi_3}  \Big) \, .
 \end{aligned}
\end{equation}

\paragraph{The $y$-directions along the $\l$-deformed spaces:}
Focusing on the $y$-components of \eqref{DilatonEinstein}, turns out that the ones surviving are along the diagonal 
directions $(y_1 y_1)$ and $(y_2 y_2)$ resulting at
\begin{equation}
 \begin{split}
  \label{Constr3AdS3}
    \nabla^2_g & \big(  A + \chi_1  \big)
  + \partial \big( A + \chi_1 \big) \cdot \partial \big(   3 \psi + \chi_1 + \chi_2 + \chi_3 + \chi_4  \big)
  - \frac{e^{- 2 \chi_1}}{\l^2_+}
  + \frac{e^{- 2 \chi_2}}{\l^2_-}
  \\
  & = - \frac{e^{- 6 \psi}}{4 k^2} \Big(  c^2_1 e^{ - 2 \chi_1 - 2 \chi_4} - c^2_2 e^{ - 2 \chi_2 - 2 \chi_3}  \Big)
 \end{split}
\end{equation}
and
\begin{equation}
 \begin{split}
  \label{Constr4AdS3}
    \nabla^2_g & \big(  A + \chi_2  \big)
  + \partial \big( A + \chi_2 \big) \cdot \partial \big(   3 \psi + \chi_1 + \chi_2 + \chi_3 + \chi_4  \big)
  + \frac{e^{- 2 \chi_1}}{\l^2_+}
  - \frac{e^{- 2 \chi_2}}{\l^2_-}
  \\
  & = \frac{e^{- 6 \psi}}{4 k^2} \Big(  c^2_1 e^{ - 2 \chi_1 - 2 \chi_4} - c^2_2 e^{ - 2 \chi_2 - 2 \chi_3}  \Big)\ ,
 \end{split}
\end{equation}
respectively.

\paragraph{The $z$-directions along the $\l$-deformed spaces:} As  before, the non-vanishing components of the Einstein equations \eqref{DilatonEinstein} are along $(z_1 z_1)$ and $(z_2 z_2)$ 
leading to
\be
 \begin{split}
  \label{Constr5AdS3}
   \nabla^2_g & \big(  A + \chi_3  \big)
  + \partial \big( A + \chi_3 \big) \cdot \partial \big(   3 \psi + \chi_1 + \chi_2 + \chi_3 + \chi_4  \big)
  + \frac{e^{- 2 \chi_3}}{\l^2_+}
  - \frac{e^{- 2 \chi_4}}{\l^2_-}
  \\
  & = \frac{e^{- 6 \psi}}{4 k^2} \Big(  c^2_1 e^{ - 2 \chi_1 - 2 \chi_4} - c^2_2 e^{ - 2 \chi_2 - 2 \chi_3}  \Big)
 \end{split}
\ee
and
\be
 \begin{split}
  \label{Constr6AdS3}
    \nabla^2_g & \big(  A + \chi_4  \big)
  + \partial \big( A + \chi_4 \big) \cdot \partial \big(   3 \psi + \chi_1 + \chi_2 + \chi_3 + \chi_4  \big)
  - \frac{e^{- 2 \chi_3}}{\l^2_+}
  + \frac{e^{- 2 \chi_4}}{\l^2_-}
  \\
  & = - \frac{e^{- 6 \psi}}{4 k^2} \Big(  c^2_1 e^{ - 2 \chi_1 - 2 \chi_4} - c^2_2 e^{ - 2 \chi_2 - 2 \chi_3}  \Big)\ ,
 \end{split}
\ee
respectively.

\paragraph{The mixed directions:}
There is also a number of mixed components that are non-trivial. These are along the $(\m y)$- and $(\m z)$-directions and they give 
rise to the following first order equations, respectively
\begin{equation}
 \partial_{\m} \big(  2 A + \chi_1 + \chi_2  \big) = 0 \, , \qquad\qquad
 \partial_{\m} \big(  2 A + \chi_3 + \chi_4  \big) = 0 \, ,
\end{equation}
integrated to
\begin{equation}
 \label{Constr7AdS3}
 2 A + \chi_1 + \chi_2 = 0 \, , \qquad\qquad
 2 A + \chi_3 + \chi_4 = 0 \, ,
\end{equation}
where the integration constants were fixed by requiring consistency with the background values \eqref{BackgrValsAdS3scalars}.

\no
The constraints \eqref{Constr7AdS3} tell us that from the six scalars only four of them are independent. In addition,
using them one can easily see that \eqref{Constr3AdS3} and \eqref{Constr4AdS3} are equivalent and similarly for \eqref{Constr5AdS3} and \eqref{Constr6AdS3}.
Hence,  we remain with the metric $g_{\m\n}$ and, by making a specific choice among the scalars, with  $A , \psi , \chi_1 , \chi_3$. The independent set of equations now will be \eqref{EinsteinReductionAdS3} together with \eqref{Constr1AdS3}, \eqref{Constr2AdS3}, \eqref{Constr3AdS3} and \eqref{Constr5AdS3}.

\subsection{A change of frame and the stability analysis}

Our equations can be further simplified in a different frame metric given by 
\begin{equation}
 \label{MetricRescalingAdS3}
 g_{\m\n} = e^{8 A - 6 \psi} \gn_{\m\n} \, .
\end{equation}
In this frame, the equations of motion for the rescaled metric $\gn_{\m\n}$ and the scalars $A , \psi , \chi_1 , \chi_3$ can be derived from an action of the form \eqref{ActionDdimensions} where now $D = 3$ and the scalars are encoded in a four-vector $X = (A , \psi , \chi_1 , \chi_3)$. The matrix $\g_{ij}$ is
\begin{equation}
 \g_{ij} = \begin{pmatrix}
               32 & -12 & 2 & 2
               \\
               -12 & 12 & 0 & 0
               \\
               2 & 0 & 2 & 0
               \\
               2 & 0 & 0 & 2
              \end{pmatrix} 
\end{equation}
and the potential $V(X)$ reads
\begin{equation}
 \begin{aligned}
  V(X)  = &  - 6 \, e^{8 A - 8 \psi}
  - 2 \, e^{8 A - 6 \psi} \, \bigg( \frac{e^{- 2 \chi_1}}{\l^2_+}
  + \frac{e^{4 A + 2 \chi_1}}{\l^2_-}
  - \frac{e^{- 2 \chi_3}}{\l^2_+}
  - \frac{e^{4 A + 2 \chi_3}}{\l^2_-} \bigg)
  \\
  & + \frac{e^{12 A - 12 \psi}}{2 k^2} \Big(  c^2_1 e^{ 2 \chi_3 - 2 \chi_1} + c^2_2 e^{ 2 \chi_1 - 2 \chi_3}  \Big) \, .
 \end{aligned}
\end{equation}
The vacuum of appendix \ref{AdS3} corresponds to the background values \eqref{BackgrValsAdS3metric} and \eqref{BackgrValsAdS3scalars}. Hence, the background value for $\gn_{\m\n}$ is related to that of $g_{\m\n}$ in \eqref{BackgrValsAdS3metric} via
\begin{equation}
 \bar{\gn}_{\m\n} = \Big(  \frac{2}{\ell} \Big)^3 \bar{g}_{\m\n} \, 
\end{equation}
and using \eqn{SolutionGravityScalars} amounts to an $AdS_3$ space with radius
\begin{equation}
 L = \frac{4}{\ell^2} \, .
\end{equation}
In order to proceed with the stability analysis we need the eigenvalues of the mass matrix square $M^2$ which is a $4\times 4 $ matrix. 
The expressions are quite complicated and not illuminating for general values of $\l$. 
Nevertheless, for $\l=0$ they become tractable so that we examine this case first. 
The result we obtain give us an insight of what to expect in general. 

\paragraph{The undeformed case with $\l=0:$}
The matrix $M^2$ is
\begin{equation}
 M^2 = \frac{\ell^4}{2} \begin{pmatrix}
                                    0 & 0 & 0 & 0
                                    \\
                                    - 1& 1 & 0 & 0
                                    \\
                                    - \frac{1}{\hat{\ell}} & 0 & \frac{1}{2} - \frac{1}{\hat{\ell}} & - \frac{1}{2}
                                    \\
                                    \frac{1}{\hat{\ell}} & 0 & - \frac{1}{2} & \frac{1}{2} + \frac{1}{\hat{\ell}}
                                    \end{pmatrix} \, ,
\end{equation}
with eigenvalues
\begin{equation}
\label{3.24}
 d_1 = 0 \, , \quad
 d_2 = \frac{\ell^4}{4} \Big(  1 + \sqrt{1 + \frac{4}{\hat{\ell}^2}} \Big) \, , \quad
 d_3 = \frac{\ell^4}{2} \, , \quad
 d_4 = \frac{\ell^4}{4} \Big(  1 - \sqrt{1 + \frac{4}{\hat{\ell}^2}} \Big) \, ,
\end{equation}
where 
\be 
 \hat\ell := k \ell\ .
\ee
From \eqn{3.24} and \eqn{massdim} we find the associated scaling dimensions
\be
 \begin{split}
  & \D^+_1 = 2 \, , \qquad \D^-_1 = 0 \, , \qquad\,\,\,\, 
     \D^\pm_2 = 1 \pm \sqrt{5 + 4 \sqrt{1 + \frac{4}{\hat{\ell}^2}}} \, ,
  \\
  & \D^+_3 = 4 \, , \qquad \D^-_3 = - 2 \, , \qquad 
     \D^\pm_4 = 1 \pm \sqrt{5 - 4 \sqrt{1 + \frac{4}{\hat{\ell}^2}}} \, .
 \end{split}
\ee
All of them, apart from $\D^\pm_4$, are manifestly real. To ensure reality of $\D^\pm_4$ we demand that
\begin{equation}
 \hat{\ell} \geqslant \frac{8}{3} \, .
\end{equation}
Therefore the radius of the $AdS_3$ requires a minimum value so that stability is not excluded, even though classically $\hat \ell >0$.

\paragraph{The general case:} Classically, reality of the supergravity solution requires a minimum value for the $AdS_3$ scale that is 
\be
\label{clabound}
\hat \ell \geqslant {4\l \ov 1-\l^2}\ . 
\ee
Our findings above for $\l=0$ suggest that stability may require a stricter bound than \eqref{clabound}, which is explicitly evaluated below. 

\no
To proceed we define the matrix 
\be
\label{Bads3}
B = \mathbb{1} + {16\ov \ell^4} M^2\ , 
\ee
with eigenvalues written in terms of $d_i$ (the eigenvalues of $M^2$) as 
\begin{equation}
 b_i= 1 + \frac{16}{\ell^4} d_i \geqslant 0 \, , \qquad i = 1 , \ldots , 4 \, .
\end{equation}
The positivity is required for stability according to  \eqn{BFbound}.
The characteristic polynomial of the matrix $B$ turns out to be 
\be
\begin{split}
& p_4(s) = (s-1) p_3(s)\ ,
\\
&  p_3(s) = s^3 - 19 s^2 + \bigg(99 - 64 {1 + 18 \l^2 + \l^4\ov \hat \ell^2(1-\l^2)^2}\bigg) s - 
\bigg(81 - 192 {3 + 22 \l^2 + 3\l^4\ov \hat \ell^2(1-\l^2)^2}\bigg)\ . 
\end{split}
\ee
From the factorisation of $p_4(s)$ clearly one eigenvalue of $B$ is unity, say $b_1=1$. The constant term of the polynomial $p_3(s)$ coming with a minus sign equals the product $b_2 b_3 b_4$.\footnote{
We use the fact that the polynomial $p_3(s)$ can also be written as
\[
 p_3(s) = (s - b_2) (s - b_3) (s - b_4) = s^3 - (b_2 + b_3 + b_4) s^2 + (b_2 b_3 + b_2 b_4 + b_3 b_4) s - b_2 b_3 b_4 \, .
\]
} In the desired scenario of stability all of the eigenvalues $(b_2, b_3, b_4)$ must be non-negative, so must be their product. This tells us that a necessary but not sufficient condition for stability is that $\hat{\ell}$ satisfies the inequality
\begin{equation} 
\label{CriticalCurvev2}
\hat{\ell} \geqslant \frac{8}{3 \, \sqrt{3}} \, \frac{\sqrt{\big.3+22 \, \lambda^2 + 3 \, \lambda^4}}{1-\lambda^2} \, , 
\end{equation}
which is clearly a stricter bound than that in \eqn{clabound}. This guarantees that $b_2 b_3 b_4 \geqslant 0$,
but not the positivity of each eigenvalue separately. There is always a possibility that one eigenvalue is positive and two negative. However this can not be true in our case. In order to show this we assume, without loss of generality, that $b_2 \geqslant 0$ and $b_3,b_4<0$. From the coefficient of the quadratic term in $p_3(s)$ we have that $b_2+b_3+b_4=19$, which together with our assumption implies that $b_2 > 19$. In addition, it can be shown that the coefficient of the linear term in $p_3(s)$ is positive for all values of $\l$. Therefore,  $b_2 b_3 + b_2 b_4 + b_3 b_4>0$. Trading $b_4$ from the aforementioned sum we have that
$-(b_2-19)(b_2+b_3) > b_3^2 > 0$. This is true for $b_2+b_3<0$ yielding that $19=b_2+b_3+b_4< b_4$. The latter contradicts to our initial assumption where $b_4 < 0$. Hence we conclude that whenever $\hat{\ell}$ satisfies \eqref{CriticalCurvev2} all eigenvalues $b_i$, $i=1,2,3,4$ are non-negative which according to \eqn{BFbound} ensures the reality of the scaling dimensions. 

\begin{figure}[h!]
 \begin{center}
  \begin{tabular}{cc}
   \includegraphics[width=0.44\textwidth]{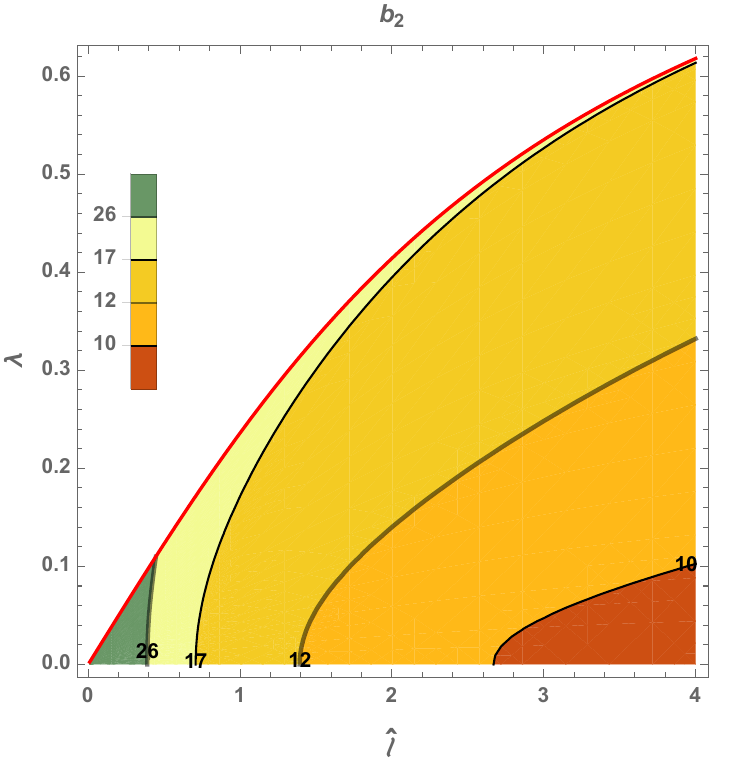}&
   \includegraphics[width=0.44\textwidth]{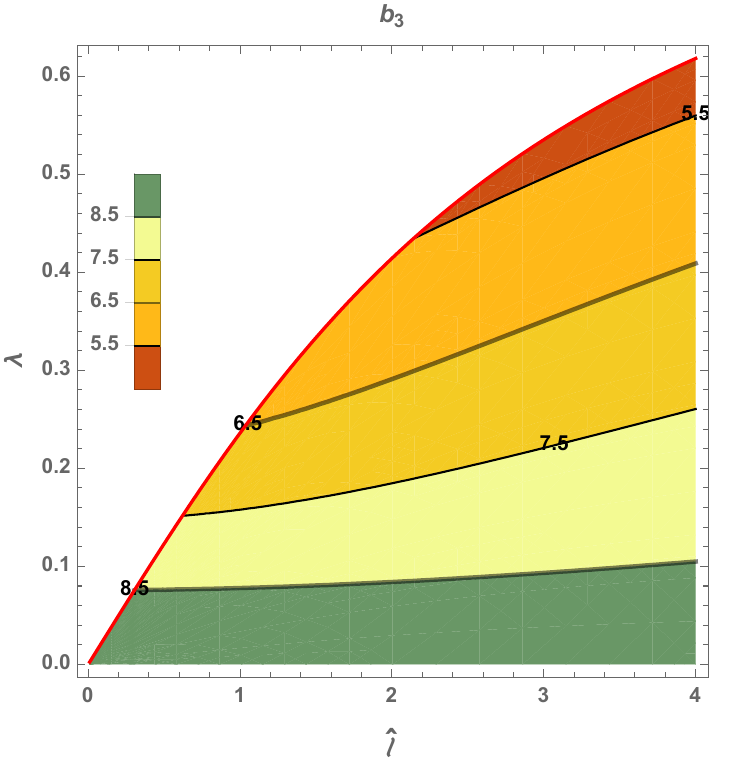}
  \end{tabular}
 \caption{\small The eigenvalues $b_2$ (left) and $b_3$ (right) of the matrix $B$ in eq. \eqref{Bads3} are constant along the contours denoted by dark lines. In the coloured areas, the values of $b_2$ and $b_3$ are in between the values attached to the contours. The curve in red parametrized by the equality in eq. \eqref{clabound} defines the allowed region of the parameter space spanned by $\l$ and $\hat{\ell}$.}
 \label{BFb2b3}
 \end{center}
\end{figure}

\begin{figure}[h!]
 \begin{center}
   \includegraphics[width=0.44\textwidth]{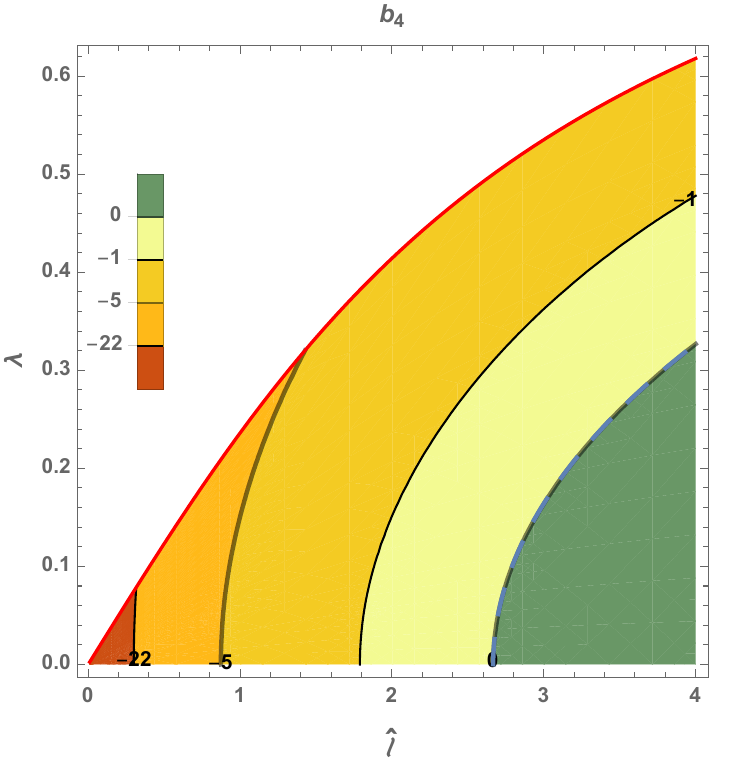}
 \caption{\small The eigenvalue $b_4$ as a function of $\l$ and $\hat{\ell}$. Here, 
 the classically allowed region in the parameter space is divided by a critical curve (dashed line) along which $b_4 = 0$. This curve is given by the equality in eq. \eqn{CriticalCurvev2}. The area on the left of this contour is where instability occurs since $b_4$ is always negative. On the right of the dashed line (denoted by green colour) $b_4$ is always positive.}
 \label{BFb4}
 \end{center}
\end{figure}

\no
The analysis above is also illustrated in 
 figures \ref{BFb2b3} and \ref{BFb4}.
 In figure \ref{BFb2b3} we plot the eigenvalues $b_2$ and $b_3$ of the matrix $B$ as a function of $\l$ and $\hat{\ell}$ parameters. The latter are confined between the horizontal axis and the curve in red, which is defined by the equality in \eqref{clabound}. In this domain the   eigenvalues  $b_2$ and $b_3$ are 
 positive, as it can be seen from the two contour plots, and thus they are not associated to unstable modes. 
The case of the eigenvalue $b_4$ is shown in figure \ref{BFb4}. There exists a critical curve (dashed line), parametrized by the 
equality in \eqref{CriticalCurvev2}, on which $b_4 = 0$. Therefore, the allowed region for the parameters $\l$ and $\hat{\ell}$ 
is divided into two sub-regions, one that sits between the red and dashed lines with $b_4 < 0$ and one that sits on the right of 
the dashed line with $b_4 > 0$. Clearly, one should disregard the area of the parameter space with $b_4 < 0$, 
where the instability of the mode associated to the eigenvalue $b_4$ occurs.


%
%

\section{The $AdS_4$ solutions}
\label{AdS4solution}

In this section we perform a stability analysis  of a class of type-IIB solutions with geometry $AdS_4 \times \cN_4 \times CS^2_\l$ where 
the four-dimensional space $\cN_4$ can be any of the following  spaces
$(S^4 , H_4 , T^4)$. All of the aforementioned cases are 
summarised in appendix \ref{AdS4} and have only one free parameter, that is $\l$, 
in addition to the level $k$ of the undeformed CFTs.
When the four-dimensional space is either the sphere $S^4$ or 
the hyperboloid $H_4$ we strictly have $\l\neq 0$.
When ${\cN_4} = T^4$, we can 
construct one background with $\l\geqslant 0$ and one with $\l=0$ but with an additional free parameter.  
The latter case shows signs of stability.

\subsection{The reduction ansatz}
\label{AnsatzAdS4}

We will adopt a reduction ansatz that fits all the solutions that are mentioned in appendix \ref{AdS4}. 
The reduction takes place on the $\l$-deformed  $CS^2_\l$ and the four-dimensional spaces $\cN_4$. 
Hence, for the metric we have the ansatz 
\begin{equation}
 \label{AnsatzAdS4metric}
 d\hat{s}^2 = e^{2 A} \Big[  ds^2_{\cM_4} + e^{2 \psi} ds^2_{\cN_4} + e^{2 \phi_y} \Big(  \l^2_+ \, e^{2 \chi_1} \, dy^2_1 + \l^2_- \, e^{2 \chi_2} \, dy^2_2  \Big) \Big] \, ,
\end{equation}
where $\cM_4$ is a four-dimensional space with metric
\begin{equation}
 ds^2_{\cM_4} = g_{\m\n} \, dx^\m \, dx^\n
\end{equation}
and $ds^2_{\cN_4}$ can be any of the line elements in \eqref{N4metric}. The scalars $A, \psi, \chi_1, \chi_2$ are taken to depend only on the coordinates $x^\m$ of $\cM_4$. For the NS three-form and the dilaton we consider the ansatz
\begin{equation}
 \label{AnsatzAdS4NS}
 \widehat{H}_3 = 0 \, , \qquad \widehat{\Phi} (x,y) = 4 A(x) + \phi_y(y) \, ,
\end{equation}
with $\phi_y(y)$ given in \eqref{CS2l}. For the RR sector we assume that
\be
 \begin{split}
  & \widehat{F}_1 = c_1 \, \l_+ \, dy_1 + c_2 \, \l_- \, dy_2 \, ,
  \qquad
  \widehat{F}_3 = 0 \, ,
  \\
  & \widehat{F}_5 = e^{- 4 \psi} \textrm{Vol}(\cM_4) \wedge \big(  c_3 e^{\chi_1 - \chi_2} \, \l_+ \, dy_1 + c_4 \, e^{\chi_2 - \chi_1} \, \l_- \, dy_2 \big) 
  \\
  &
  \qq + \textrm{Vol}(\cN_4) \wedge \big(  c_4 \, \l_+ \, dy_1 - c_3 \, \l_- \, dy_2 \big) \, ,
 \end{split}
\ee
where $\textrm{Vol}(\cM_4)$ is the volume form on $\cM_4$ and $\textrm{Vol}(\cN_4)$ is given by the corresponding expression 
in \eqref{VolN4}.

\no
The solutions in appendix \ref{AdS4} can be obtained after setting $\cM_4$ to be an $AdS_4$ space with metric
\begin{equation}
 \label{BackgrValsAdS4metric}
 ds^2_{AdS_4} = \bar{g}_{\m\n} \, dx^\m \, dx^\n = \frac{3}{\ell_1} \bigg( r^2\eta_{\a\b} dx^\a dx^\b + \frac{dr^2}{r^2} \bigg) \, ,
\end{equation}
normalised as $\bar{R}_{\m\n} = - \ell_1 \, \bar{g}_{\m\n}$ and by setting the scalars to
\begin{equation}
 \label{BackgrValsAdS4scalars}
 \bar A = \bar \psi = \bar \chi_1 =\bar \chi_2 = 0\ .
\end{equation}
The constants $c_1 , \ldots , c_4$ and $\ell_1 , \ell_2$ take values according to the six different solutions 
 given by \eqref{solIAdS4}, \eqref{solIIAdS4}, \eqref{solIIIAdS4}, \eqref{solIVAdS4}, \eqref{solVAdS4} and \eqref{solVIAdS4}.

\subsection{The equations of motion}

We determine the equations of motion for the scalars $A , \psi , \chi_1 , \chi_2$ by inserting the ansatz of the subsection \ref{AnsatzAdS4} into the equations of the type-IIB supergravity presented in appendix \ref{IIB}. Similarly to the analysis for the $AdS_3$ solution, we observe that the equations for the form fields \eqref{BianchiIIB} and \eqref{FluxesIIB} are trivially satisfied. Thus we only have to deal with the dilaton and Einstein equations \eqref{DilatonEinstein}. 

\no
Tensors constructed below and all contractions are performed with respect to the metric $g_{\m\n}$ on $\cM_4$.

\paragraph{The dilaton equation:}
The dilaton equation \eqref{DilatonEinstein} reduces to
\be
 \label{DilatonReductionAdS4}
  \begin{split}
   R &+ 4 \, \e \, \ell_2 \, e^{- 2 \psi}
  - 2 \nabla^2 \big(  A + 4 \psi + \chi_1 + \chi_2 \big)
 \\
 &  - \big(  4 \partial \psi + \partial \chi_1 + \partial \chi_2 \big)^2
  - 4 \big( \partial \psi \big)^2   - 8 \big(  \partial A \big)^2 
 - \big( \partial \chi_1 \big)^2 - \big( \partial \chi_2 \big)^2 
  \\
  &
  - 2 \partial \big(  4 \psi + \chi_1 + \chi_2 \big) \cdot \partial A
  + \frac{2}{\l^2_+} e^{- 2 \chi_1}
  + \frac{2}{\l^2_-} e^{- 2 \chi_2} = 0 \, ,
 \end{split}
\ee
The $\e$ symbol takes the values $0 , \pm1$ according to \eqref{N4scaling}.

\paragraph{The directions along $\cM_4$:}
If we restrict ourselves to the components of the Einstein equations \eqref{DilatonEinstein} along the $\cM_4$ directions we get
\be
 \label{EinsteinReductionAdS4}
 \begin{split}
   R_{\m\n} & - \nabla_\m \nabla_\n \big(  4 \psi + \chi_1 + \chi_2 \big)
  - 4 \del_\m \psi \del_\n \psi
  - \del_\m \chi_1 \del_\n \chi_1
  - \del_\m \chi_2 \del_\n \chi_2
  \\ 
  & 
  - 8 \del_\m A \del_\n A
  - g_{\m\n} \nabla^2 A
   - g_{\m\n} \partial \big(  4 \psi + \chi_1 + \chi_2 \big) \cdot \partial A
  \\
  & + g_{\m\n} \bigg(  \frac{e^{8 A}}{4} \Big(  c^2_1 e^{- 2 \chi_1} + c^2_2 e^{- 2 \chi_2} \Big)
  + \frac{e^{- 8 \psi}}{4} \Big(  c^2_3 e^{- 2 \chi_2} + c^2_4 e^{- 2 \chi_1} \Big)  \bigg) = 0 \, .
 \end{split}
\ee

\no
Taking the trace of the previous equation and combining it with \eqref{DilatonReductionAdS4} in order to eliminate the Ricci scalar we find
that
\be
\begin{split}
\label{Constr1AdS4}
   \nabla^2 &\big(  4 \psi + \chi_1 + \chi_2 - 2 A \big)
  - 2 \partial \big(  4 \psi + \chi_1 + \chi_2 \big) \cdot \partial A
  \\
  &+ \big(  4 \partial \psi + \partial \chi_1 + \partial \chi_2 \big)^2
- 4 \, \e \, \ell_2 \, e^{- 2 \psi}
  - \frac{2}{\l^2_+} e^{- 2 \chi_1}
  - \frac{2}{\l^2_-} e^{- 2 \chi_2}
  \\
  &
  + e^{8 A} \Big(  c^2_1 e^{- 2 \chi_1} + c^2_2 e^{- 2 \chi_2} \Big)
  + e^{- 8 \psi} \Big(  c^2_3 e^{- 2 \chi_2} + c^2_4 e^{- 2 \chi_1} \Big) = 0 \, .
 \end{split}
\ee
 We will use this equation instead of the equivalent one in \eqn{DilatonReductionAdS4}.

\paragraph{The directions along  $\cN_4$:}
The only non-trivial components of the Einstein equations along the $\cN_4$ directions are the diagonal ones. They 
all lead to the same equation given by
\begin{equation}
 \label{Constr2AdS4}
   \begin{aligned}
  & \nabla^2 \big(  A + \psi \big)
  + \partial \big(  4 \psi + \chi_1 + \chi_2 \big) \cdot \partial \big(  A + \psi \big)
  \\
  & \quad - \e \, \ell_2 \, e^{- 2 \psi}
  - \frac{e^{8 A}}{4} \Big(  c^2_1 e^{- 2 \chi_1} + c^2_2 e^{- 2 \chi_2} \Big)
  + \frac{e^{- 8 \psi}}{4} \Big(  c^2_3 e^{- 2 \chi_2} + c^2_4 e^{- 2 \chi_1} \Big) = 0 \, .
 \end{aligned}
\end{equation}

\paragraph{The directions along the $\l$-deformed space:}
Focusing on the $y$-components of \eqref{DilatonEinstein} we find that  the diagonal ones, i.e. the $(y_1 y_1)$ and $(y_2 y_2)$, 
contribute, respectively, as
\be
\begin{split}
 \label{Constr3AdS4}
  &  \nabla^2  \big(  A + \chi_1 \big)
  + \partial \big(  4 \psi + \chi_1 + \chi_2 \big) \cdot \partial \big(  A + \chi_1 \big)
  \\
  & \quad   - \frac{e^{- 2 \chi_1}}{\l^2_+}
  + \frac{e^{- 2 \chi_2}}{\l^2_-}
  + \frac{e^{8 A}}{4} \Big(  c^2_1 e^{- 2 \chi_1} - c^2_2 e^{- 2 \chi_2} \Big)
  - \frac{e^{- 8 \psi}}{4} \Big(  c^2_3 e^{- 2 \chi_2} - c^2_4 e^{- 2 \chi_1} \Big) = 0 \, 
\end{split}
\ee
and
\be
\begin{split}
 \label{Constr4AdS4}
   & \nabla^2  \big(  A + \chi_2 \big)
  + \partial \big(  4 \psi + \chi_1 + \chi_2 \big) \cdot \partial \big(  A + \chi_2 \big)
  \\
  & \ + \frac{e^{- 2 \chi_1}}{\l^2_+}
  - \frac{e^{- 2 \chi_2}}{\l^2_-}
  - \frac{e^{8 A}}{4} \Big(  c^2_1 e^{- 2 \chi_1} - c^2_2 e^{- 2 \chi_2} \Big)
  + \frac{e^{- 8 \psi}}{4} \Big(  c^2_3 e^{- 2 \chi_2} - c^2_4 e^{- 2 \chi_1} \Big) = 0 \, .
\end{split}
\ee
The off-diagonal component $(y_1 y_2)$ provides the condition 
\begin{equation}
 \label{Constr5AdS4}
 c_1 \, c_2 \, e^{8 A} - c_3 \, c_4 \, e^{- 8 \psi} = 0 \, .
\end{equation}
This is obviously satisfied for the solutions I - IV in appendix \ref{AdS4}. However, for V \& VI \eqref{Constr5AdS4} 
may be considered as an extra constraint on the scalars $A$ and $\psi$. Hence, the various solutions have to be treated
separately as far as the stability analysis is concerned. 

\paragraph{The mixed directions:}
Finally, non-trivial components of the Einstein equations \eqref{DilatonEinstein} along the mixed $(\m y)$ directions give rise to the 
first order equation
\begin{equation}
 \label{FirstOrderConstr}
 \partial_{\m} \big(  2 A + \chi_1 + \chi_2  \big) = 0 \, .
\end{equation}
This is integrated to
\begin{equation}
 \label{Constr6AdS4}
 2 A + \chi_1 + \chi_2 = 0 \, ,
\end{equation}
where the integration constant is fixed by the background values \eqref{BackgrValsAdS4scalars}. Due to this constraint, \eqn{Constr3AdS4} and \eqn{Constr4AdS4} are equivalent.

\no
From the considerations above it is clear that for the solutions I - IV, 
we are left with equations \eqref{EinsteinReductionAdS4}, \eqref{Constr1AdS4}, \eqref{Constr2AdS4} and \eqref{Constr3AdS4} 
for the metric $g_{\m\n}$ on $\cM_4$ and the three scalars $\psi , \chi_1$ and  $\chi_2$. 
The scalar $A$ should be substituted everywhere using \eqref{Constr6AdS4}.
For the solutions V \& VI we have in addition the constraint  \eqref{Constr5AdS4} which leaves the possibility for a 
reduced number of scalars as we will see in detail.


\subsection{A change of frame and the stability analysis}

\begin{figure}[h]
 \begin{center}
  \begin{tabular}{cc}
   \includegraphics[width=0.47\textwidth]{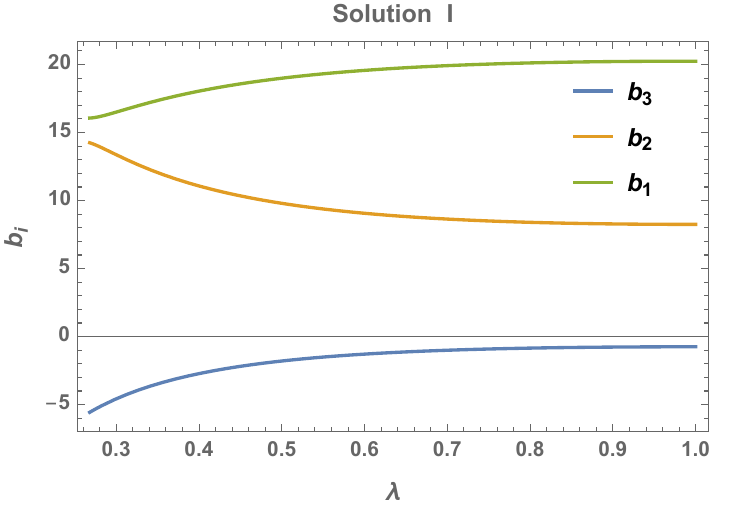}&
   \includegraphics[width=0.47\textwidth]{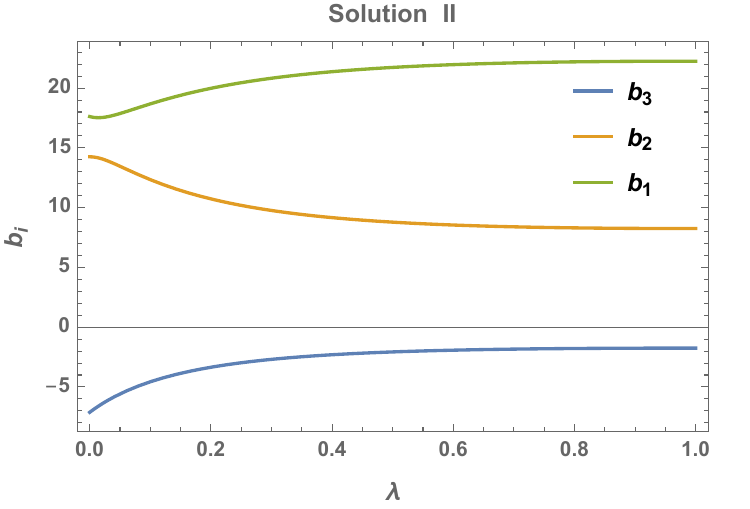}
   \\[5pt]
   \includegraphics[width=0.47\textwidth]{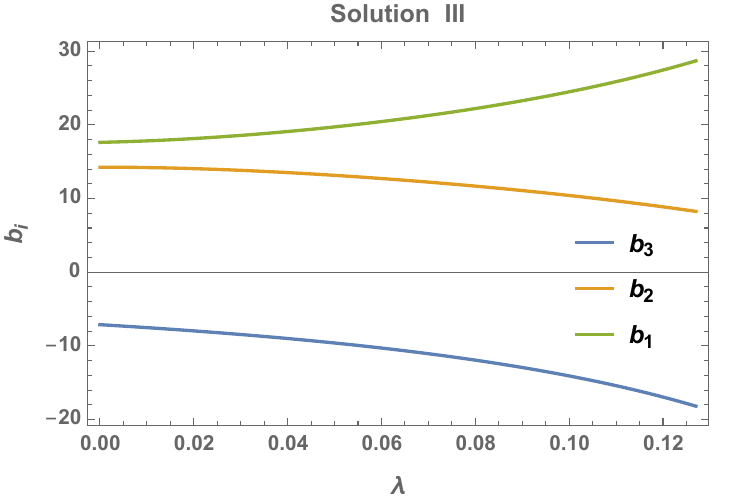}&
   \includegraphics[width=0.47\textwidth]{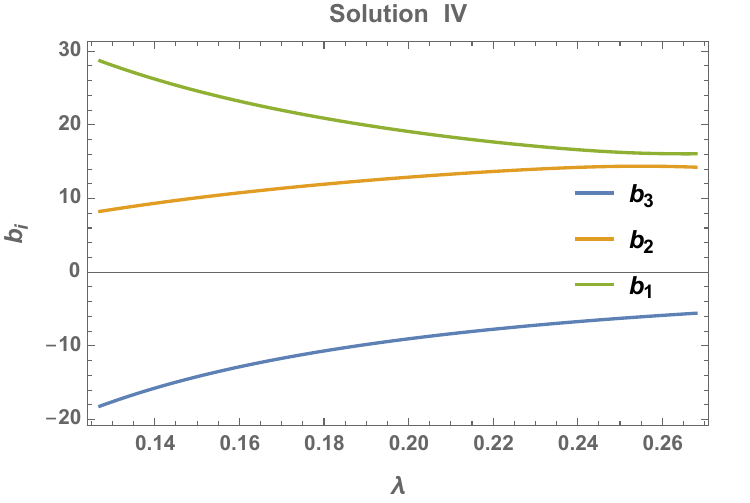}
  \end{tabular}
 \caption{\small The eigenvalues $b_i$ in eq. \eqref{BFfunctionAdS4} as a function of 
 the deformation parameter $\l$. The upper left plot refers to the solution I, the upper right to the solution II, 
 the lower left to the solution III and the lower right to the solution IV. In all cases there is a negative eigenvalue, namely the $b_3$, signaling the existence of an unstable mode.}
 \label{BFAdS4}
 \end{center}
\end{figure}

\paragraph{The solutions I-IV:} The equations  can be further simplified if we change the metric frame as
\begin{equation}
 \label{MetricRescaling1AdS4}
 g_{\m\n} = e^{- 4 \psi - \chi_1 - \chi_2} \gn_{\m\n} \, .
\end{equation}
Then, it turns out that the equations of motion for the metric $\gn_{\m\n}$ and the scalars $ \psi , \chi_1 , \chi_2$ can be obtained by an action of the form \eqref{ActionDdimensions} where now $D = 4$ with the vector for the scalars being 
$X = (\psi , \chi_1 , \chi_2)$. In addition, the matrix $\g_{ij}$ is
\begin{equation}
 \label{gamma1AdS4}
 \g_{ij} = \frac{1}{2} \begin{pmatrix}
                                 24 & 4 & 4
                                 \\
                                 4 & 7 & 5
                                 \\
                                 4 & 5 & 7
                               \end{pmatrix}\ ,
\end{equation}
whereas  the potential $V(X)$ is
\be
 \label{V1AdS4}
\begin{split} 
  V(X)  = & 2 \, e^{- 4 \psi - \chi_1 - \chi_2} \bigg(  c^2_1 \frac{e^{- 6\chi_1 - 4 \chi_2}}{4}
  + c^2_2 \frac{e^{- 4 \chi_1 - 6 \chi_2}}{4}
  + c^2_3 \frac{e^{- 8 \psi - 2 \chi_2}}{4}
  + c^2_4 \frac{e^{- 8 \psi - 2 \chi_1}}{4}
  \\
 & - \frac{e^{- 2 \chi_1}}{\l^2_+}
 - \frac{e^{- 2 \chi_2}}{\l^2_-}
 - 2 \, \e \, \ell_2 \, e^{- 2 \psi}  \bigg) \, .
 \end{split}
 \ee
The vacua associated with the solutions I-IV of appendix \ref{AdS4} correspond to \eqref{BackgrValsAdS4metric} and \eqref{BackgrValsAdS4scalars}. Notice that, although we changed the frame according to the eq. \eqref{MetricRescaling1AdS4} the background values for $\gn_{\m\n}$ and $g_{\m\n}$ are the same, i.e.
\begin{equation}
 \bar{\gn}_{\m\n} = \bar{g}_{\m\n} \, ,
\end{equation}
and that amounts to an $AdS_4$ of radius
\begin{equation}
 L = \sqrt{\frac{3}{\ell_1}} \, .
\end{equation}
\no
To study the stability of the fluctuations around the vacua I-IV 
we define, similarly to  \eqn{Bads3}, the matrix 
\be
\label{Bads4}
B = \frac{9}{4} \mathbb{1} + \frac{3}{\ell_1 }M^2\ , 
\ee
with eigenvalues 
\begin{equation}
 \label{BFfunctionAdS4}
 b_i = \frac{9}{4} + \frac{3}{\ell_1} d_i \geqslant 0 \, ,
\end{equation}
where $d_i$ are the eigenvalues of the matrix $M^2$ and positivity is according to  \eqn{BFbound}.
We have computed for each one of the four solutions the characteristic polynomial of matrix $B$ which is cubic in order. 
We will not present their explicit expressions but nevertheless, by examining the constant and quadratic pieces we easily conclude 
that one eigenvalue is negative and two positive for all  allowed values of $\l$. 
These are illustrated in figure \ref{BFAdS4} where we plot the eigenvalues $b_i$ as a function of $\l$ for each one of the four solutions.

\paragraph{The solution V:}
This solution can be separated in two distinct cases. In the first one, the deformation parameter $\lambda$ takes the specific value 
$\l = 2 - \sqrt{3}$ and the constraint \eqref{Constr5AdS4} is trivially satisfied. As a result this case can be seen as a continuation 
of the previous one, in the sense that we can use the same four-dimensional action to obtain the equations of motion for the 
metric and the scalars. Following this line of reasoning, the corresponding mass matrix reads
\begin{equation}
 M^2 = \frac{1}{2 \sqrt{3} k}
  \begin{pmatrix}
   16 & -1 & 1
   \\
   -16 & - 5 & 5
   \\
   16 & 19 & 13
  \end{pmatrix} \, .
\end{equation}
From this we compute the eigenvalues of the matrix $B$ through \eqref{BFfunctionAdS4} which are
\begin{equation}
 b_1 = \frac{57}{4} \, , \qquad b_2 = \frac{21}{4} + 3 \sqrt{13} \, , \qquad b_3 = \frac{21}{4} - 3 \sqrt{13} \, .
\end{equation}
Obviously the mode associated to the eigenvalue $b_3$ is unstable as it violates the BF bound.

When $\l \in [0 , 2 - \sqrt{3})$ the case of V solution differs significantly from those of I-IV, 
since now there is an extra constraint for the scalars. This comes from \eqref{Constr5AdS4} which now implies
\begin{equation}
 \label{Apsi}
 A + \psi = 0 \, .
\end{equation}
We can use this and \eqn{Constr6AdS4} to obtain $\psi=\ha (\chi_1+\chi_2)$. Thus we stay only with the metric 
on $\cM_4$ and the scalars $\chi_1 , \chi_2$.
Again we can go to the convenient Einstein frame letting
\begin{equation}
 \label{MetricRescaling2AdS4}
 g_{\m\n} = e^{- 3 (\chi_1 + \chi_2)} \gn_{\m\n} \, .
\end{equation}
In the new frame, the equations of motion for the metric $\gn_{\m\n}$ and the scalars $\chi_1 , \chi_2$ can be derived from the four-dimensional analogue of the action \eqref{ActionDdimensions} with $X = (\chi_1 , \chi_2)$, the matrix $\g_{ij}$ being
\begin{equation}
 \label{gamma2AdS4}
 \g_{ij} = \frac{1}{2} \begin{pmatrix}
                                 17 & 15
                                 \\
                                 15 & 17
                               \end{pmatrix}
\end{equation}
and the potential $V(X)$
\begin{equation}
\begin{split}
 V(X) = & e^{- 7 \chi_1 - 7 \chi_2} \Big[  \Big(  \frac{\n}{4} - \m  \Big) e^{- 2 \chi_1} +  \Big(  \frac{\n}{4} + \m  \Big) e^{- 2 \chi_2} \Big]
 \\
 &
  - \frac{2}{\l^2_+} e^{- 5 \chi_1 - 3 \chi_2}
  - \frac{2}{\l^2_-} e^{- 3 \chi_1 - 5 \chi_2} \, .
\end{split}
\end{equation}
Using \eqref{M2matrix} we can construct the $2 \times 2$ mass matrix
\begin{equation}
 M^2 =
  \frac{1}{4} \begin{pmatrix}
                    \n - 17 \m & 3 \n - 15 \m
                     \\
                    3 \n + 15 \m & \n + 17 \m
                  \end{pmatrix}
\end{equation}
and from its eigenvalues we evaluate $b_1$ and $b_2$ via \eqref{BFfunctionAdS4} giving
\begin{equation}
 b_1 = \frac{21}{4} + \frac{3}{\n} \sqrt{9 \n^2 + 64 \m^2} \, , \qquad
 b_2 = \frac{21}{4} - \frac{3}{\n} \sqrt{9 \n^2 + 64 \m^2} \, .
\end{equation}
The mode associated with $b_2$ always violates the BF bound as it is negative for all  the 
values of $\l$ in the range $[0 , 2 - \sqrt{3})$.


\paragraph{The solution VI:} In this example a more interesting structure appears. Initially we observe that the constraint 
\eqref{Constr5AdS4} is trivially satisfied when the constant $c_1$, i.e. the free parameter entering in the solution VI through 
the coefficients of the RR fields $F_1$ and $F_5$ in \eqref{solVIAdS4}) takes the value
$c_1 = \pm \sqrt{\frac{2}{k}}$. When this happens one can safely use the four-dimensional version of the action \eqref{ActionDdimensions} with $\g_{ij}$ given in \eqref{gamma1AdS4} and $V(X)$ in \eqref{V1AdS4}. As usual, the matrix $M^2$ is computed from eq. \eqref{M2matrix} and has the form
\begin{equation}
  M^2 = \frac{1}{2 k}
  \begin{pmatrix}
   8 & -1 & 1
    \\
   -8 & 7 & 9
    \\
   8 & 3 & -3
  \end{pmatrix} \, .
\end{equation}
The eigenvalues of this matrix and those for the matrix $B$ defined in \eqref{Bads4} are related via \eqref{BFfunctionAdS4}. 
We find that
\begin{equation}
 b_1 = \frac{57}{4} \, , \qquad b_2 = \frac{21}{4} + 3 \sqrt{17} \, , \qquad b_3 = \frac{21}{4} - 3 \sqrt{17} \, .
\end{equation}
Since $b_3 < 0$ we conclude that the corresponding mode is unstable.

Another option for the parameter $c_1$ is to allow $|c_1| < \sqrt{\frac{2}{k}}$. In this case 
the constraint \eqref{Constr5AdS4} implies that $A$ and $\psi$ are related as in eq. \eqref{Apsi}. 
Substituting the relation between $A$ and $\psi$ in \eqref{Constr2AdS4} we arrive to the following new constraint
\begin{equation}
\label{kjdsch}
 \bigg( \frac{1}{k} - c^2_1 \bigg) \Big( e^{- 2 \chi_1} - e^{- 2 \chi_2} \Big) = 0 \, .
\end{equation}
This is satisfied in two ways: Either when $c_1 = \pm \sqrt{\frac{1}{k}}$, which, as explained below \eqref{BoundAdS4l0}, 
is identical with the $\l = 0$ limit of the solution V and thus it has one unstable mode or,
if $c_1 \ne \pm \sqrt{\frac{1}{k}}$, by taking $\chi_1 = \chi_2 := \chi$. Therefore, we end up with a four-dimensional system of gravity with one scalar described by the action
\begin{equation}
 S = \frac{1}{2 \k^2_4} \int d^4 x \sqrt{|\gn|} \bigg(  \Rn - 32 \big(   \partial \chi  \big)^2 - V(\chi) \bigg)
 \quad {\rm with} \quad
 V(\chi) = \frac{2}{k} e^{- 16 \chi}
  - \frac{4}{k} e^{- 8 \chi} \, .
\end{equation}
The scalar $\chi$ has mass squared $M^2 = \frac{4}{k}$ and thus the BF bound is not violated.
This is the only case in the whole analysis of the $\lambda$-deformed gravity solutions with an $AdS_4$ factor that an 
instability is not present. 

Summarising, we arrive at the following conclusion: 
The stability analysis of the $\lambda$-deformed type-IIB backgrounds with an $AdS_4$ factor uncovered 
an unstable mode for all solutions I-V, and only for the
solution VI there are {\it islands} of potential stability. Precisely, in the case of the  $AdS_4 \times T^4 \times CS^2$
background ($\lambda$=0), with the values of the free parameter $c_1$ satisfying $|c_1| < \sqrt{2/k}$ and excluding the values $ c_1=\pm \sqrt{1/k}$, our stability analysis
did not  detect any sign of instability. Technically, the feature that distinguishes the case VI from the other five (I-V) is that one is forced,
cf. \eqn{kjdsch}, to take the two scalars equal, i.e. $\chi_1=\chi_2$. In turn, this gives rise to a positive mass term around the minimum of the single scalar potential. 


\section{The $AdS_6$ solutions}
\label{AdS6solution}

Following the same line of the previous sections we examine the stability of  a class of type-IIB solutions with geometry 
$AdS_6 \times \cN_2 \times CS^2_\l$, where now the two-dimensional space $\cN_2$ is one of the following spaces 
$(S^2 , H_2 , T^2)$. More details on these  solutions 
can be found in appendix \ref{AdS6}. Likewise to the preceding section, $\l$ is the only independent parameter,
apart from the level $k$ of the undeformed CFT.

\subsection{The reduction ansatz}
\label{AnsatzAdS6}

A reduction ansatz for the metric accommodating the three solutions of appendix B.3 has the form
\begin{equation}
 \label{AnsatzAdS6metric}
 d\hat{s}^2 = e^{2 A} \Big[  ds^2_{\cM_6} + e^{2 \psi} ds^2_{\cN_2} + e^{2 \phi_y} \Big(  \l^2_+ \, e^{2 \chi_1} \, dy^2_1 + \l^2_- \, e^{2 \chi_2} \, dy^2_2  \Big) \Big] \, ,
\end{equation}
where $\cM_6$ is a six-dimensional space with metric
\begin{equation}
 ds^2_{\cM_6} = g_{\m\n} \, dx^\m \, dx^\n
\end{equation}
and $ds^2_{\cN_2}$ can be any of the line elements in \eqref{N2metric}. Also, the scalars $A, \psi, \chi_1, \chi_2$ are taken to depend exclusively on the coordinates $x^\m$ of $\cM_6$. For the NS three-form and the dilaton we consider
\begin{equation}
 \label{AnsatzAdS6NS}
 \widehat{H}_3 = 0 \, , \qquad \widehat{\Phi} (x,y) = 4 A(x) + \phi_y(y) \, ,
\end{equation}
where $\phi_y(y)$ is given by \eqref{CS2l} and for the RR sector we assume that 
\begin{equation}
 \begin{aligned}
  \widehat{F}_1 = c_1 \, \l_+ \, dy_1 + c_2 \, \l_- \, dy_2 \, ,
  \quad
  \widehat{F}_3 = \textrm{Vol}(\cN_2) \wedge \big(   c_3 \, \l_+ \, dy_1 + c_4 \, \l_- \, dy_2  \big) \, ,
  \quad
  \widehat{F}_5 = 0 \, ,
 \end{aligned}
\end{equation}
with $\textrm{Vol}(\cN_2)$ given by the appropriate expression in \eqref{VolN2}.

\no
The vacua of appendix \ref{AdS6} can be obtained after taking $\cM_6$ to be an $AdS_6$ with
\begin{equation}
 \label{BackgrValsAdS6metric}
 ds^2_{AdS_6} = \bar{g}_{\m\n} \, dx^\m \, dx^\n = \frac{5}{\ell_1} \Big(   r^2 dx^\a dx^\b + \frac{dr^2}{r^2} \Big) \, ,
\end{equation}
normalised as $\bar{R}_{\m\n} = - \ell_1 \, \bar{g}_{\m\n}$ and by setting the scalars to
\begin{equation}
 \label{BackgrValsAdS6scalars}
 \bar A =  \bar \psi = \bar \chi_1 =  \bar\chi_2 = 0\, .
\end{equation}
The constants $c_1 , \ldots , c_4$ and $\ell_1 , \ell_2$ take values according to the three different solutions  
in \eqref{solIAdS6}, \eqref{solIIAdS6} and \eqref{solIIIAdS6}.

\subsection{The equations of motion}

We determine the equations of motion for the scalars $A , \psi , \chi_1 , \chi_2$ by plugging our ansatz 
into the equations of type-IIB supergravity in appendix \ref{IIB}. In the same manner as the examples containing $AdS_3$ and $AdS_4$ factors, the form field equations  \eqref{BianchiIIB} and \eqref{FluxesIIB} are trivially satisfied. Thus we only have to work out the dilaton and the Einstein equations \eqref{DilatonEinstein}.

\no
 Tensors constructed below and all contractions are performed with respect to the metric $g_{\m\n}$ on $\cM_6$.

\paragraph{The dilaton equation:}The dilaton equation \eqref{DilatonEinstein} reduces to
\be
\begin{split}
 \label{DilatonReductionAdS6}
   R & + 2 \, \e \, \ell_2 \, e^{- 2 \psi}
  - 2 \nabla^2 \big(  A + 2 \psi + \chi_1 + \chi_2 \big)
  - \big(  2 \partial \psi + \partial \chi_1 + \partial \chi_2 \big)^2
  - 2 \big( \partial \psi \big)^2
  \\
  &
    - \big( \partial \chi_1 \big)^2
  - \big( \partial \chi_2 \big)^2
 - 8 \big(  \partial A \big)^2
  - 2 \partial \big(  2 \psi + \chi_1 + \chi_2 \big) \cdot \partial A
  \\
  &
  + \frac{2}{\l^2_+} e^{- 2 \chi_1}
  + \frac{2}{\l^2_-} e^{- 2 \chi_2} = 0 \, ,
 \end{split}
\ee
The $\e$ symbol takes the values $0 , \pm1$ according to \eqref{N2scaling}.

\paragraph{The directions along $\cM_6$:}
The components of the Einstein equations \eqref{DilatonEinstein} along $\cM_6$ give
\be
\begin{split}
 \label{EinsteinReductionAdS6}
   R_{\m\n} & - \nabla_\m \nabla_\n \big(  2 \psi + \chi_1 + \chi_2 \big)
  - 2 \del_\m \psi \del_\n \psi
  - \del_\m \chi_1 \del_\n \chi_1
  - \del_\m \chi_2 \del_\n \chi_2
  \\
  &
  - 8 \del_\m A \del_\n A
  - g_{\m\n} \big( \nabla^2 A +  \partial \big(  2 \psi + \chi_1 + \chi_2 \big) \cdot \partial A\big)
   \\ 
   &
  + g_{\m\n} \Bigg[  \frac{e^{8 A}}{4} \Big(  c^2_1 e^{- 2 \chi_1} + c^2_2 e^{- 2 \chi_2} \Big)
  + \frac{e^{4 A - 4 \psi}}{4} \Big(  c^2_3 e^{- 2 \chi_1} + c^2_4 e^{- 2 \chi_2} \Big)  \Bigg] = 0 \, .
 \end{split}
\ee

\no
The trace of the last equation combined with \eqref{DilatonReductionAdS6} serves to eliminate the Ricci scalar from it giving
\begin{equation}
\label{Constr1AdS6}
  \begin{aligned}
   \nabla^2 \big(  2 \psi &+ \chi_1 + \chi_2 - 4 A \big)
  - 4 \partial \big(  2 \psi + \chi_1 + \chi_2 \big) \cdot \partial A
  + \big(  2 \partial \psi + \partial \chi_1 + \partial \chi_2 \big)^2
  \\
  & - 2 \, \e \, \ell_2 \, e^{- 2 \psi}
  - \frac{2}{\l^2_+} e^{- 2 \chi_1}
  - \frac{2}{\l^2_-} e^{- 2 \chi_2}
  + \frac{3}{2} e^{8 A} \Big(  c^2_1 e^{- 2 \chi_1} + c^2_2 e^{- 2 \chi_2} \Big)
  \\
&   + \frac{3}{2} e^{4 A - 4 \psi} \Big(  c^2_3 e^{- 2 \chi_1} + c^2_4 e^{- 2 \chi_2} \Big) = 0 \, .
 \end{aligned}
\end{equation}
It is convenient to use this equation instead of the equivalent one in \eqn{DilatonReductionAdS6}.

\paragraph{The directions along $\cN_2$:}
Along these directions the only non-trivial components of Einstein's equations
are the diagonal ones leading to the same expression
\be
 \label{Constr2AdS6}
   \begin{split}
  & \nabla^2 \big(  A + \psi \big)
  + \partial \big(  2 \psi + \chi_1 + \chi_2 \big) \cdot \partial \big(  A + \psi \big)
  \\
  &\qq - \e \, \ell_2 \, e^{- 2 \psi}
  - \frac{e^{8 A}}{4} \Big(  c^2_1 e^{- 2 \chi_1} + c^2_2 e^{- 2 \chi_2} \Big)
  + \frac{e^{4 A - 4 \psi}}{4} \Big(  c^2_3 e^{- 2 \chi_1} + c^2_4 e^{- 2 \chi_2} \Big) = 0 \, .
 \end{split}
\ee

\paragraph{The directions along the $\l$-deformed space:} The diagonal  $y$-components of \eqref{DilatonEinstein}, 
i.e. the $(y_1 y_1)$ and $(y_2 y_2)$, give
\be
 \label{Constr3AdS6}
 \begin{split}
  \nabla^2 \big(  A &  + \chi_1 \big)
  + \partial \big(  2 \psi + \chi_1 + \chi_2 \big) \cdot \partial \big(  A + \chi_1 \big)  - \frac{e^{- 2 \chi_1}}{\l^2_+}
  + \frac{e^{- 2 \chi_2}}{\l^2_-}
  \\
  &
  + \frac{e^{8 A}}{4} \Big(  c^2_1 e^{- 2 \chi_1} - c^2_2 e^{- 2 \chi_2} \Big)
  + \frac{e^{4 A - 4 \psi}}{4} \Big(  c^2_3 e^{- 2 \chi_1} - c^2_4 e^{- 2 \chi_2} \Big) = 0 \, 
 \end{split}
\ee
and
\be
 \label{Constr4AdS6}
 \begin{split}
   \nabla^2 \big(  A & + \chi_2 \big)
  + \partial \big(  2 \psi + \chi_1 + \chi_2 \big) \cdot \partial \big(  A + \chi_2 \big)  + \frac{e^{- 2 \chi_1}}{\l^2_+}
  - \frac{e^{- 2 \chi_2}}{\l^2_-}
  \\
  &
  - \frac{e^{8 A}}{4} \Big(  c^2_1 e^{- 2 \chi_1} - c^2_2 e^{- 2 \chi_2} \Big)
  - \frac{e^{4 A - 4 \psi}}{4} \Big(  c^2_3 e^{- 2 \chi_1} - c^2_4 e^{- 2 \chi_2} \Big) = 0 \, ,
 \end{split}
\ee
respectively. The off-diagonal component is trivially satisfied.

\paragraph{The mixed components:} The integration of the equation arising from the mixed $(\m y)$ directions provides that
\begin{equation}
 \label{Constr6AdS6}
 2 A + \chi_1 + \chi_2 = 0 \, ,
\end{equation}
in accordance with the background values \eqref{BackgrValsAdS6scalars}. This can be used to solve for $A$ and eliminate it from the equations. 
Hence, the metric $g_{\m\n}$ on $\cM_6$ and the independent scalars $\psi , \chi_1 , \chi_2$ must satisfy equations \eqref{EinsteinReductionAdS6}, \eqref{Constr1AdS6}, \eqref{Constr2AdS6} and \eqref{Constr3AdS6}, where we note that  \eqref{Constr4AdS6} is equivalent to  \eqref{Constr3AdS6}.

\subsection{A change of frame and the stability analysis}

If we further change metric frame as
\begin{equation}
 \label{MetricRescaling1AdS6}
 g_{\m\n} = e^{-  \psi - \frac{\chi_1 + \chi_2}{2}} \gn_{\m\n} \, ,
\end{equation}
then, the equations of motion for the metric and the scalars can be obtained by an action of the form \eqref{ActionDdimensions} where now $D = 6$ and the scalars are encoded into the vector $X = (\psi , \chi_1 , \chi_2)$. 
The matrix $\g_{ij}$ is
\begin{equation}
 \label{gamma1AdS6}
 \g_{ij} = \frac{1}{4} \begin{pmatrix}
                                 12 & 2 & 2
                                 \\
                                 2 & 13 & 9
                                 \\
                                 2 & 9 & 13
                               \end{pmatrix}\ .
\end{equation}
Also the potential $V(X)$ is
\be
 \label{VAdS6}
 \begin{split}
  V(X) & = 2 \, e^{- \psi - \frac{\chi_1 + \chi_2}{2}} \bigg(  c^2_1 \frac{e^{- 6\chi_1 - 4 \chi_2}}{4}
  + c^2_2 \frac{e^{- 4 \chi_1 - 6 \chi_2}}{4}
  + c^2_3 \frac{e^{- 4 \psi - 4 \chi_1 - 2 \chi_2}}{4}
  \\
 & + c^2_4 \frac{e^{- 4 \psi - 2 \chi_1 - 4 \chi_2}}{4}
  - \frac{e^{- 2 \chi_1}}{\l^2_+}
 - \frac{e^{- 2 \chi_2}}{\l^2_-}
 - \e \, \ell_2 \, e^{- 2 \psi}  \bigg) \, .
 \end{split}
\ee
We now study the linearized equations of motion for the scalars $\psi , \chi_1 , \chi_2$ around the $AdS_6$ solutions I-III of appendix \ref{AdS6}. These correspond to the background  \eqref{BackgrValsAdS6metric} and \eqref{BackgrValsAdS6scalars}. 

\no
From \eqref{MetricRescaling1AdS6} we observe that the background values for $\gn_{\m\n}$ and $g_{\m\n}$ are the same, i.e.
\begin{equation}
 \bar{\gn}_{\m\n} = \bar{g}_{\m\n} \, ,
\end{equation}
and thus $\gn_{\m\n}$ refers to an $AdS_6$ of radius
\begin{equation}
 L = \sqrt{\frac{5}{\ell_1}} \, .
\end{equation}
Following the lines of the section \ref{GravityScalars} we construct the matrix
\begin{equation}
 \label{BAdS6}
 B = \frac{25}{4} \mathbb{1} + \frac{5}{\ell_1} \, M^2 \, ,
\end{equation}
whose eigenvalues are
\begin{equation}
 b_i = \frac{25}{4} + \frac{5}{\ell_1} \, d_i \geqslant 0 \, , \qquad i = 1 , 2 , 3 \, ,
\end{equation}
with $d_i$ being the eigenvalues of the matrix $M^2$ defined in eq. \eqref{M2matrix}. Positivity of the $b_i$'s is necessary for stability. 
Below we analyse this for each solution of appendix \ref{AdS6}.

\begin{figure}[h!]
 \begin{center}
  \begin{tabular}{cc}
   \includegraphics[width=0.47\textwidth]{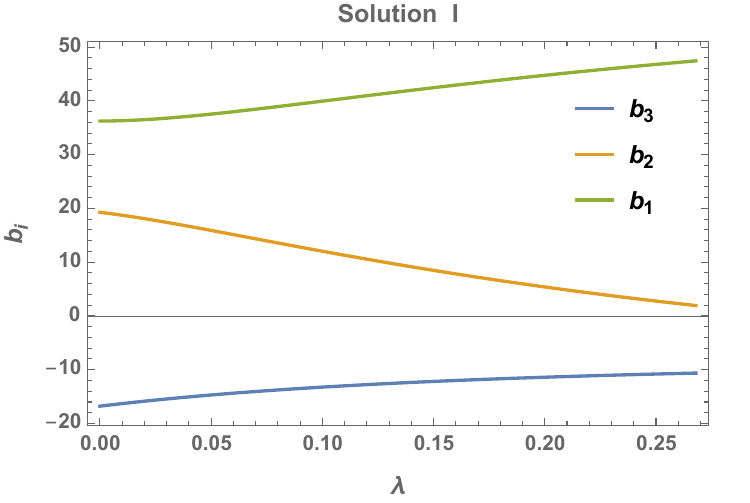}&
   \includegraphics[width=0.47\textwidth]{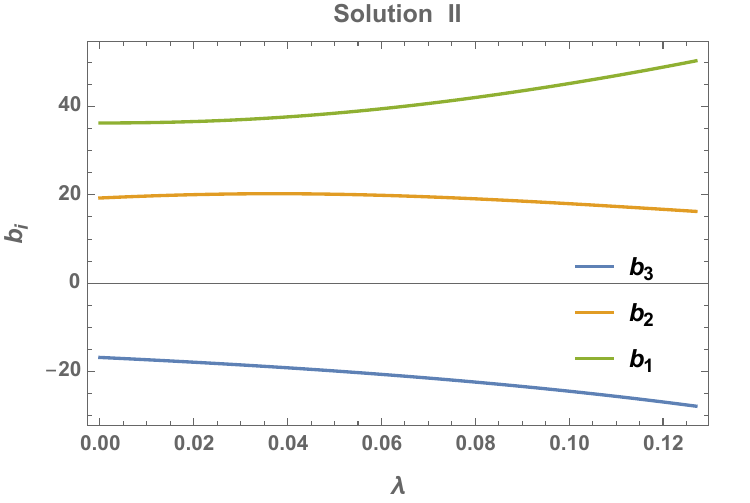}
  \end{tabular}
 \caption{\small The eigenvalues $b_i$ as a function of $\l$. The left plot corresponds to the solution I and 
 the right to the solution II. The existence of the negative eigenvalue $b_3$ in both cases suggests that the $AdS_6$ solutions I and II are unstable.}
 \label{BFAdS6}
 \end{center}
\end{figure}

\paragraph{The solutions I \& II:}
It turns out that when $\cN_2 = S^2$ or $\cN_2 = H_2$ one of the eigenvalues is negative for all the allowed values of $\l$ whereas the other two stay positive. Hence the backgrounds are unstable. This pattern is illustrated in figure \ref{BFAdS6}, where for each solution we plot the eigenvalues of $B$ matrix as a function of $\l$. Clearly, one of them,
say $b_3$, is always negative for all values of $\l$.

\paragraph{The solution III:}
This solution exists only for $\l=0$. The mass matrix $M^2$ from \eqref{M2matrix} is
\begin{equation}
 M^2 = \frac{1}{6 k}
  \begin{pmatrix}
   12 & 3 & 9
    \\
   -4 & 7 & 21
    \\
   12 & 15 & -3
  \end{pmatrix} \, .
  \ee
The corresponding matrix $B$ constructed from \eqref{BAdS6} has the eigenvalues
\begin{equation}
 b_1 = \frac{145}{4} \, , \qquad b_2 = \frac{5}{4} + 5 \sqrt{13} \, , \qquad b_3 = \frac{5}{4} - 5 \sqrt{13} \, .
\end{equation}
Obviously, since $b_3 < 0$ the related mode is unstable. 
Notice that although this solution is not the smooth $\lambda \rightarrow 0$ limit of I (one should also replace 
the sphere by the torus) the result of the stability analysis is qualitatively the same: one of the 
eigenvalues leads to an unstable mode. At this point, there is a qualitative difference compared to the $AdS_4$ result, and the reason 
(as we discussed at the end of the previous section) is that the 
$\lambda=0$ limit in the  $AdS_4$ case is accompanied with a simplification in the reduction ansatz (i.e. $\chi_1 =\chi_2 = \chi$), 
something that is absent in the $AdS_6$ case.


\section{Conclusions}
\label{conclusions}

In the current paper we study a wide class of relatively simple supergravity backgrounds
constructed in \cite{Itsios:2019izt}. 
These are non-supersymmetric solutions with an unwarped $AdS$ factor and originate from embedding 
integrable $\lambda$-deformed $\sigma$-models to type-II supergravity. 
All of these solutions contain a continuous parameter $\lambda$, taking values in a certain range between $0$ and $1$.
The absence of supersymmetry, the presence of the $AdS$ factor and the Ooguri--Vafa conjecture \cite{Ooguri:2016pdq}, as well as the potential usefulness 
within the AdS/CFT correspondence, were the mains motivations to investigate the stability of these solutions, starting with the perturbative one.

Among the different solutions constructed in \cite{Itsios:2019izt} we chose to analyse those with an 
$AdS_n$ factor of $n=3,4,6$. All of them fall into the class of solutions that the Ooguri-Vafa conjecture applies.
In the present work we examined their perturbative stability.  
Our study was based on a consistent dimensional reduction of these solutions
to a lower dimensional theory with scalars coupled to gravity and a subsequent perturbative stability analysis of the corresponding lower dimensional theory. The study of the different backgrounds revealed that most of them are perturbatively unstable except for the $\l$-deformed solution with geometry $AdS_3 \times S^3 \times CS^2_\l \times CH_{2,\l}$ and the undeformed one with geometry $AdS_4 \times T^4 \times CS^2$ (solution VI in appendix \ref{AdS4}). Below we mention the solutions resisting the characterisation unstable.

The analysis of the $AdS_3 \times S^3 \times CS^2_\l \times CH_{2,\l}$ solution uncovers an interesting structure. 
The conformal scaling factors  related to the fluctuations of the scalars are functions of the deformation parameter $\lambda$ and the combination $\hat \ell =k \ell$, where $k$ is the level of the undeformed CFT and $\ell $ is the curvature scale of  $AdS_3$, thus creating a two-dimensional parameter space to search for violation of the BF bound. As it is depicted in figure \ref{BFb2b3} two of the eigenvalues of the matrix $B$ are clearly positive and thus are not associated to unstable modes. On the other hand, the figure \ref{BFb4} shows that there is an eigenvalue which is negative in the region of the parametric space confined between the red and dashed lines and positive elsewhere. 
The area where this eigenvalue is negative must be disregarded while its complement requires further investigation.

In the case of $AdS_4 \times T^4 \times CS^2$ there is another free 
parameter that it is called $c_1$ appearing in the RR sector. It turns out that when $|c_1|< \sqrt{2/k}$ (excluding the values with
$|c_1|=1/\sqrt{k}$), our stability analysis does not detect unstable modes violating the BF bound. As such, this background serves as another candidate for a more elaborate study.

Equally important with the perturbative instabilities are the non-perturbative ones. 
Current progress in that front is coming from a novel decay channel, introduced in \cite{Bena:2020xxb},
and comes under the name brane-jet instability. In this set-up, a probe brane is placed in the background 
and the force acting on the brane has to be examined. In case this force is repulsive, the vacuum is characterised unstable. In \cite{Bena:2020xxb}, 
one of the few gravity backgrounds that challenged the Ooguri-Vafa conjecture, i.e. the 
non-supersymmetric $SO(3) \times SO(3)$ invariant $AdS_4$ vacuum of
the 4-dimensional ${\cal N} =8$  $SO(8)$ gauged supergravity \cite{Warner:1983vz}, 
was proved brane-jet unstable. In a parallel line of research, tachyonic Kaluza-Klein modes were found 
for the same background \cite{Malek:2020mlk}. Brane-jet instabilities also arise for other $AdS$ vacua \cite{Suh:2020rma}.

The fact that the perturbative analysis of the $AdS_3$ and $AdS_4$ solutions that survived leave a substantial part of the parametric space free of instabilities suggests that a brane-jet calculation could be useful to further constrain the {\it window} of potential stability. However, such a computation for the $AdS_3$ case is not conclusive since it is not easy to extract the potential generating the forces on the brane. This is due to the involved dependence of the RR potential on the internal coordinates. As such, we can not infer about the existence of possible non-perturbative instabilities based on a brane-jet argument. The same reasoning holds for the $AdS_4$ example as well.

The analysis of the current paper leaves a number of important open questions: In the $AdS_3$ case the reduction ansatz we 
considered and the subsequent perturbative analysis left an important part of the parametric space 
$(\lambda, \hat{\ell})$ to challenge the Ooguri-Vafa conjecture. Applying more powerful techniques, for instance those coming from 
exceptional field theory and calculating the full Kaluza-Klein spectrum would unambiguously decide about the presence or not 
of perturbative instabilities. The same kind of analysis could also be applied to the $AdS_4$ undeformed case and for the range of values of the parameter $c_1$ that the current analysis does not identify as violating the BF bound. 

For those solutions that the perturbative analysis does not detect instabilities, a further exploration of the 
field theory needs to be put forward by applying the standard tools of the AdS/CFT correspondence. Calculation of the Maxwell
and Page charges and their dependence on the $\lambda$ parameter,  the study of the dynamics and of the mesonic spectrum are some computations that could be done.

Consistent truncations are typically related to the presence of a symmetry, direct or hidden. For $\lambda$-deformed backgrounds such symmetries are not present, except for the undeformed cases (i.e. $\lambda=0$) which have a $U(1)$ isometry. Investigating the presence of a hidden symmetry is an interesting future direction. In addition, progress in that front would be extremely helpful in order to perform a systematic search and decide whether the lower dimensional effective theories are part of a consistent reduction of a 10-dimensional supergravity. At present we are not aware that this is the case.


\subsection*{Acknowledgments}

We would like to thank C. Nu\~nez and E. Malek for carefully reading
the preliminary draft and for sending various useful comments. We also thank K. Pilch and  K. Siampos for a correspondence and for
useful discussions.\\
The present work was co-funded by the European Union and
Greek national funds through the Operational Program "Human Resources
Development, Education and Lifelong Learning" (NSRF 2014-2020), under the
call "Supporting Researchers with an Emphasis on Young Researchers - Cycle
B" (MIS: 5047947).


\appendix

\section{Type-IIB supergravity}
\label{IIB}

In this appendix we review the field content and equations of motion of type-IIB supergravity. It is described by the following 
string frame action
\be
 \begin{split}
  \cS_{IIB} & = \frac{1}{2 \k^2_{10}} \int_{M_{10}} \Big[  e^{- 2 \Phi} \Big(  R \star \mathbb{1} + 4 d \Phi \wedge \star d \Phi - \frac{1}{2} H_3 \wedge \star H_3 \Big)
  \\
 & \qquad - \frac{1}{2} \Big(  F_1 \wedge \star F_1 + F_3 \wedge \star F_3 + \frac{1}{2} F_5 \wedge \star F_5  \Big) \Big] 
 + \cS_{\rm t.t.} \, ,
 \end{split}
\ee
where $\cS_{\rm t.t.}$ is a topological term given by
\begin{equation}
 \cS_{t.t.} = - \frac{1}{4 \k^2_{10}} \int_{M_{10}} C_4 \wedge H_3 \wedge d C_2 \, .
\end{equation}
The field content of type-IIB supergravity consists of the metric $G_{MN}$ on the ten-dimensional space $M_{10}$, a dilaton $\Phi$, a NS two-form $B_2$ whose field strength is $H_3$ and the RR potentials $C_0 , C_2 , C_4$ which give rise to the higher-rank forms $F_1 , F_3 , F_5$ through
\begin{equation}
 H_3 = d B_2 \, , \qquad 
 F_1 = d C_0 \, , \qquad 
 F_3 = d C_2 - C_0 H_3 \, , \qquad 
 F_5 = d C_4 - H_3 \wedge C_2 \, .
\end{equation}
Thus, the form fields satisfy the following Bianchi identities
\begin{equation}
 \label{BianchiIIB}
 d H_3 = 0 \, , \qquad
 d F_1 = 0 \, , \qquad
 d F_3 = H_3 \wedge F_1 \, , \qquad
 d F_5 = H_3 \wedge F_3 \, .
\end{equation}
The five-form $F_5$ is self-dual, i.e. $\star F_5 =F_5$ which is imposed by hand.

\no
The equations of motion arising from variations of the dilaton and the metric are
\be
 \begin{split}
 \label{DilatonEinstein}
  & R \star \mathbb{1} + 4 \, d \star d \Phi - 4 \, d\Phi \wedge \star d\Phi - \frac{1}{2} H_3 \wedge \star H_3 = 0 \, ,
  \\
  & R_{MN} + 2 \, \nabla_M \nabla_N \Phi - \frac{1}{4} \big(  H^2_3  \big)_{MN} = \frac{e^{2 \Phi}}{2} \bigg( \big(  F^2_1 \big)_{MN} + \frac{1}{2} \big(  F^2_3  \big)_{MN} + \frac{1}{48} \big(  F^2_5 \big)_{MN}
  \\
  & \qquad\qquad\qquad\qquad\qquad\qquad\qq - G_{MN} \Big(  \frac{1}{2} F^2_1 + \frac{1}{12} F^2_3  \Big)  \bigg) \, ,
 \end{split}
\ee
while those arising from the variations of the RR potentials are 
\be
 \begin{split}
 \label{FluxesIIB}
  & d \Big(  e^{- 2 \Phi} \star H_3 \Big) - F_1 \wedge \star F_3 - F_3 \wedge F_5 = 0 \, ,
  \\
  & d\star F_3 + H_3 \wedge F_5 = 0 \, ,
  \\
  & d\star F_1 + H_3 \wedge \star F_3 = 0\ .
 \end{split}
\ee


\section{Supergravity solutions with $AdS$ and $\l$-deformed spaces}
\label{lambda-solutions}

Here we summarise the supergravity backgrounds found in \cite{Itsios:2019izt}, whose the perturbative stability is analysed in the main text. Before doing that let us introduce the following set of parameters which appear often in this study
\begin{equation}
 \label{constants}
 \l_{\pm} = \sqrt{k \frac{1 \pm \l}{1 \mp \l}} = \frac{k}{\l_\mp} \, , \qquad \m = \frac{4 \l}{k \big(  1 - \l^2 \big)} \, , \qquad \n = \frac{4}{k} \frac{1 + \l^2}{1 - \l^2} \, .
\end{equation}
In the above, $k$ is the level of the associated CFTs we mention below, which is a positive 
number and in addition an integer in the compact case. The  deformation parameter $\l$ 
in principle takes values in the interval $[0 , 1)$. However, each solution below may impose further restrictions on the allowed values of it.

\no 
Since all solutions are based on the $\l$-deformation of the gauged WZW models corresponding to the exact coset CFTs, $SU(2)/U(1)$ and $SL(2,\mathbb{R})/SO(1,1)$, it is also useful to introduce the metrics and the dilatons for these spaces.
For the $\l$-deformed model on $SU(2)/U(1)$ we have that
\begin{equation}
 \label{CS2l}
  ds^2_{CS^2_\l} = e^{2 \phi_y} \Big(  \l^2_+ \, dy^2_1 + \l^2_- \, dy^2_2  \Big) \, ,
  \qquad\qquad
  \phi_y(y) = - \frac{1}{2} \ln \big(  1 - y^2_1 - y^2_2 \big) \, ,
\end{equation}
where the coordinates $(y_1 , y_2)$ are restricted inside the unit disc $y^2_1 + y^2_2 < 1$. This space will be denoted as $CS^2_\l$.
Similarly, the $\l$-deformed model on $SL(2,\mathbb{R})/SO(1,1)$ is
\begin{equation}
 \label{CH2l}
  ds^2_{CH_{2,\l}} = e^{2 \phi_z} \Big(  \l^2_+ \, dz^2_1 + \l^2_- \, dz^2_2  \Big) \, ,
  \qquad\qquad
  \phi_z(z) = - \frac{1}{2} \ln \big(  z^2_1 + z^2_2 - 1 \big) \, ,
\end{equation}
where now the coordinates $(z_1 , z_2)$ lie outside the unit disc, i.e. $z^2_1 + z^2_2 > 1$. This space will be denoted as $CH_{2,\l}$.

\subsection{The $AdS_3 \times S^3 \times CS^2_\l \times CH_{2,\l}$}
\label{AdS3}

The NS sector of this solution contains a metric that takes the form
\begin{equation}
\begin{split}
 \label{solAdS3metric}
 ds^2 = \frac{2}{\ell} & \Big(  - r^2 dt^2 + r^2 dx^2 + \frac{dr^2}{r^2} 
  + d\th^2_1 + \sin^2\th_1 \, d\th^2_2 + \sin^2\th_1 \sin^2\th_2 \, d\th^2_3 \Big)
 \\
 &  + ds^2_{CS^2_\l}+ ds^2_{CH_{2,\l}} \, ,
\end{split}
\end{equation}
where $\ell$ is a constant and the line elements for $CS^2_\l$ and $CH_{2,\l}$ are given in \eqref{CS2l} and \eqref{CH2l}, respectively. There is also a dilaton whose expression is
\begin{equation}
 \Phi(y,z) = \phi_y(y) + \phi_z(z) \, ,
\end{equation}
where the functions $\phi_y(y)$ and $\phi_z(z)$ are given in \eqref{CS2l} and \eqref{CH2l}. The NS two-form $B_2$ is trivial and so is its field strength $H_3$.

\no
The above is supported by a RR sector whose content is
\begin{equation}
 \begin{aligned}
  F_1 & = 0 \, , \qquad F_3 = 0 \, ,
  \\
  F_5 & = 2 \, k \, \Big(  \frac{2}{\ell} \Big)^{\frac{3}{2}} \, dz_1 \wedge dy_2 \wedge \Bigg(  \sqrt{\frac{\ell - \m}{2}} \textrm{Vol}(AdS_3) + \sqrt{\frac{\ell + \m}{2}} \textrm{Vol}(S^3) \Bigg)
  \\
  & - 2 \, k \, \Big(  \frac{2}{\ell} \Big)^{\frac{3}{2}} \, dz_2 \wedge dy_1 \wedge \Bigg(  \sqrt{\frac{\ell + \m}{2}} \textrm{Vol}(AdS_3) + \sqrt{\frac{\ell - \m}{2}} \textrm{Vol}(S^3)  \Bigg) \, ,
 \end{aligned}
\end{equation}
where we have defined the volume forms on $AdS_3$ and $S^3$ as
\begin{equation}
 \label{VolsAdS3S3}
 \textrm{Vol}(AdS_3) = r \, dt \wedge dx \wedge dr \, , \qquad 
 \textrm{Vol}(S^3) = \sin^2\th_1 \, \sin\th_2 \, d\th_1 \wedge d\th_2 \wedge d\th_3 \, .
\end{equation}
In order for the solution to be real one has to require that $\ell \geqslant \m$.

\subsection{The $AdS_4 \times \cN_4 \times CS^2_\l$}
\label{AdS4}

Another class of type-IIB solutions with $AdS$ and $\l$-deformed factors have the line element
\begin{equation}
 \label{solAdS4metric}
 ds^2 = \frac{3}{\ell_1} \Big(  - r^2 dt^2 + r^2 dx^2_1 + r^2 dx^2_2 + \frac{dr^2}{r^2} \Big)
  + ds^2_{\cN_4}
  + ds^2_{CS^2_\l} \, ,
\end{equation}
where $\ell_1$ is a positive constant, $ds^2_{CS^2_\l}$ is given in \eqref{CS2l} and $\cN_4$ can be any of the four-dimensional spaces in the list $(S^4 , H_4 , T^4)$.  The line elements are explicitly given by
\begin{equation}
 \begin{aligned}
  \label{N4metric}
   ds^2_{S^4} &= \frac{3}{\ell_2} \Big(  d\th^2_1 + \sin^2\th_1 \, d\th^2_2 + \sin^2\th_1 \sin^2\th_2 \, d\th^2_3 + \sin^2\th_1 \sin^2\th_2 \sin^2\th_3 \, d\th^2_4   \Big) \, ,
   \\
   ds^2_{H_4} &= \frac{3}{\ell_2} \Big(  d\th^2_1 + \sinh^2\th_1 \, d\th^2_2 + \sinh^2\th_1 \sin^2\th_2 \, d\th^2_3 + \sinh^2\th_1 \sin^2\th_2 \sin^2\th_3 \, d\th^2_4   \Big) \, ,
   \\
   ds^2_{T^4} &= d\th^2_1 + d\th^2_2 + d\th^2_3 + d\th^2_4 \, .
 \end{aligned}
\end{equation}

The space $\cN_4$ is normalised such that
\begin{equation}
 \label{N4scaling}
 R^{(\cN_4)}_{MN} = \e \, \ell_2 \, g^{(\cN_4)}_{MN} \, , \qquad \textrm{with} \qquad
 \e = \begin{cases}
         + 1 & \textrm{for} \quad \cN_4 = S^4
         \\
           \,\,\,\, 0  & \textrm{for} \quad \cN_4 = T^4
         \\
         - 1 & \textrm{for} \quad \cN_4 = H_4
        \end{cases} \, ,
\end{equation}
where $\ell_2$ is also a positive constant. Moreover, the NS two-form is zero and the dilaton depends on the $y$-coordinates of the deformed space as
\begin{equation}
 \Phi = \phi_y(y) \, ,
\end{equation}
with $\phi_y(y)$ being the function in \eqref{CS2l}.

\no
The RR sector can be written in a universal fashion as
\ba
  && F_1 = c_1 \, \l_+ \, dy_1 + c_2 \, \l_- \, dy_2 \, ,
  \qquad F_3 = 0 \, ,
\\
  && F_5 = \textrm{Vol}(AdS_4) \wedge \big(  c_3 \, \l_+ \, dy_1 + c_4 \, \l_- \, dy_2 \big) + \textrm{Vol}(\cN_4) \wedge \big(  c_4 \, \l_+ \, dy_1 - c_3 \, \l_- \, dy_2 \big) \, ,
 \nonumber
 \ea
where the volume forms are
\begin{equation}
 \label{VolN4}
 \begin{aligned}
  & \textrm{Vol}(AdS_4) = \frac{9}{\ell^2_1} \, r^2 \, dt \wedge dx_1 \wedge dx_2 \wedge dr \, ,
   \\
  & \textrm{Vol}(\cN_4) =
  \begin{cases}
   \frac{9}{\ell^2_2} \, \sin^3\th_1 \, \sin^2\th_2 \, \sin\th_3 \, d\th_1 \wedge d\th_2 \wedge d\th_3 \wedge d\th_4 & \textrm{for} \quad \cN_4 = S^4
    \\
   d\th_1 \wedge d\th_2 \wedge d\th_3 \wedge d\th_4 & \textrm{for} \quad \cN_4 = T^4               
    \\
   \frac{9}{\ell^2_2} \, \sinh^3\th_1 \, \sin^2\th_2 \, \sin\th_3 \, d\th_1 \wedge d\th_2 \wedge d\th_3 \wedge d\th_4 & \textrm{for} \quad \cN_4 = H_4\ .
  \end{cases}
  \, 
 \end{aligned}
\end{equation}

\no
The parameters $c_1 , \ldots , c_4$ and $\ell_1 , \ell_2$ are chosen so that 
the type-IIB supergravity equations are solved. Below we list each one of them, separately.

\paragraph{Solution I:}
This is a solution where $\cN_4 = S^4$, with the various constants being
\begin{equation}
 \label{solIAdS4}
 c_1 = c_4 = 0 \, , \quad  c_2 = s_2 \sqrt{\frac{\n}{2}} \, , \quad c_3 = s_3 \sqrt{\frac{8 \m - \n}{2}} \, , \quad \ell_1 = \m \, , \qquad \ell_2 = \m - \frac{\n}{4} \, ,
\end{equation}
and $s_{2 , 3} = \pm 1$. The solution is well defined when $c^2_1 , c^2_2 , c^2_3 , c^2_4 \geqslant 0$ and $\ell_1 , \ell_2 > 0$. This means that $\l$ is bounded as
\begin{equation}
 2 - \sqrt{3} < \l < 1 \, .
\end{equation}

\paragraph{Solution II:}
A second solution of the same class, i.e. $\cN_4 = S^4$ is
\begin{equation}
 \label{solIIAdS4}
 c_2 = c_4 = 0 \, , \quad c_1 = s_1 \sqrt{\frac{\n}{2}} \, , \quad c_3 = s_3 \sqrt{\frac{8 \m + \n}{2}} \, , \quad \ell_1 = \frac{\n}{4} + \m \, , \qquad \ell_2 = \m \, ,
\end{equation}
with $s_{1,3} = \pm 1$.  The constant $\ell_2$ vanishes at $\l = 0$ and the rest are well behaved for $\l \in [0 , 1)$. Thus requiring $\ell_2 > 0$ implies
\begin{equation}
 0 < \l < 1 \, .
\end{equation}

\paragraph{Solution III:}
A third solution which now belongs to the class $\cN_4 = H_4$ is
\begin{equation}
 \label{solIIIAdS4}
 c_1 = c_3 = 0 \, , \quad c_2 = s_2 \sqrt{\frac{\n}{2}} \, , \quad c_4 = s_4 \sqrt{\frac{\n - 8 \m}{2}} \, , \quad \ell_1 = \frac{\n}{4} - \m \, , \quad \ell_2 = \m \, ,
\end{equation}
with $s_{2,4} = \pm 1$. This behaves well when $\l$ is restricted in the interval below
\begin{equation}
 0 < \l \leqslant 4 - \sqrt{15} \, .
\end{equation}

\paragraph{Solution IV:}
Another solution of this class, i.e. with $\cN_4 = H_4$ is
\begin{equation}
 \label{solIVAdS4}
 c_1 = c_4 = 0 \, , \quad c_2 = s_2 \sqrt{\frac{\n}{2}} \, , \quad c_3 = s_3 \sqrt{\frac{8 \m - \n}{2}} \, , \quad \ell_1 = \m \, , \quad \ell_2 = \frac{\n}{4} - \m \, ,
\end{equation}
with $s_{2,3} = \pm 1$. In this case the allowed values of $\l$ are
\begin{equation}
 4 - \sqrt{15} \leqslant \l < 2 - \sqrt{3} \, .
\end{equation}

\paragraph{Solution V:}
A distinct class of backgrounds is when $\cN_4 = T^4$. There are two solutions belonging to this class, the deformed one with $\l \geqslant 0$ and the undeformed one with $\l = 0$. For the deformed one we have that
\begin{equation}
 \label{solVAdS4}
 \begin{aligned}
  & c_1 = s_1 \, \sqrt{\frac{\n}{4} - \m} \, , \quad c_2 = s_2 \, \sqrt{\frac{\n}{4} + \m} \, ,
  \quad
  c_3 = s_3 \, \sqrt{\frac{\n}{4} + \m} \, , \quad c_4 = s_4 \, \sqrt{\frac{\n}{4} - \m} \, ,
  \\
  & \ell_1 = \frac{\n}{4} \, ,
 \end{aligned}
\end{equation}
with $s_{1 , 2 , 3 , 4} = \pm 1$ satisfying the condition $s_1 \, s_2 = s_3 \, s_4$. The solution is real when
\begin{equation}
 0 \leqslant \l \leqslant 2 - \sqrt{3} \, .
\end{equation}

\paragraph{Solution VI:}
The last solution of the class $\cN_4 = T^4$ is one with $\l$ being strictly zero and thus it has topology $AdS_4 \times T^4 \times CS^2$. This solution has $c_1$ as a free parameter, The rest of them are
\begin{equation}
 \label{solVIAdS4}
 c_2 = s_2 \, \sqrt{\frac{2}{k} - c^2_1} \, ,
 \qquad
 c_3 = s_3 \, c_1 \, ,
 \qquad
 c_4 = s_4 \, \sqrt{\frac{2}{k} - c^2_1} \, ,
 \qquad
 \ell_1 = \frac{1}{k} \, ,
\end{equation}
with $s_{2 , 3 , 4} = \pm 1$ satisfying $s_2 = s_3 \, s_4$. Reality implies that
\begin{equation}
 \label{BoundAdS4l0}
 - \sqrt{\frac{2}{k}} \leqslant c_1 \leqslant \sqrt{\frac{2}{k}} \, .
\end{equation}
Notice that when $c_1 = \pm \sqrt{\frac{1}{k}}$ one obtains the $\l = 0$ limit of \eqref{solVAdS4}. On the other hand, when $c_1 = 0$ one finds the $\l = 0$ limit of \eqref{solIIIAdS4} with $H_4$ replaced by $T^4$, while whenever $c_1 = \pm \sqrt{\frac{2}{k}}$ one recovers the $\l = 0$ limit of \eqref{solIIAdS4} with $S^4$ replaced by $T^4$.

\subsection{The $AdS_6 \times \cN_2 \times CS^2_\l$}
\label{AdS6}

The last class of solutions of interest in the present work, is that with geometry containing an $AdS_6$ part and the $\l$-deformed 
space $CS^2_\l$. The NS-sector of these backgrounds includes a metric
\begin{equation}
 \label{solAdS6metric}
 ds^2 = \frac{5}{\ell_1} \Big(  - r^2 dt^2 + r^2 dx^2_1 + r^2 dx^2_2 + r^2 dx^2_3 + r^2 dx^2_4  + \frac{dr^2}{r^2} \Big)
  + ds^2_{\cN_2}
  + ds^2_{CS^2_\l} \, ,
\end{equation}
where $\ell_1$ is a positive constant, $ds^2_{CS^2_\l}$ is given in \eqref{CS2l} and $\cN_2$ can be any of the two-dimensional spaces in the list $(S^2 , H_2 , T^2)$ with line elements
\begin{equation}
 \label{N2metric}
 ds^2_{\cN_2} = 
 \begin{cases}
  \frac{1}{\ell_2} \Big(  d\th^2_1 + \sin^2\th_1 \, d\th^2_2  \Big) & \textrm{for} \quad \cN_2 = S^2
   \\
   d\th^2_1 + d\th^2_2 & \textrm{for} \quad \cN_2 = T^2
   \\
   \frac{1}{\ell_2} \Big(  d\th^2_1 + \sinh^2\th_1 \, d\th^2_2   \Big) & \textrm{for} \quad \cN_2 = H_2\ .
  \end{cases}
\end{equation}
The space $\cN_2$ is normalised such that
\begin{equation}
 \label{N2scaling}
 R^{(\cN_2)}_{MN} = \e \, \ell_2 \, g^{(\cN_2)}_{MN} \, , \qquad \textrm{with} \qquad
 \e = \begin{cases}
         + 1 & \textrm{for} \quad \cN_2 = S^2
         \\
           \,\,\,\, 0  & \textrm{for} \quad \cN_2 = T^2
         \\
         - 1 & \textrm{for} \quad \cN_2 = H_2\ ,
        \end{cases} \, 
\end{equation}
where $\ell_2$ is also a positive constant. The rest of the NS fields are
\begin{equation}
 H_3 = 0 \, , \qquad \Phi = \phi_y(y) \, ,
\end{equation}
with $\phi_y(y)$ being the function in \eqref{CS2l}.

\no
The field content of the RR-sector is
\begin{equation}
 F_1 = c_1 \, \l_+ \, dy_1 + c_2 \, \l_- \, dy_2 \, ,
  \quad
 F_3 = \textrm{Vol}(\cN_2) \wedge \big(   c_3 \, \l_+ \, dy_1 + c_4 \, \l_- \, dy_2  \big) \, ,
  \quad
 F_5 = 0 \, ,
\end{equation}
where the volume form $\textrm{Vol}(\cN_2)$ can have any of the following expressions
\begin{equation}
 \label{VolN2}
 \textrm{Vol}(\cN_2) =
  \begin{cases}
   \frac{1}{\ell_2} \, \sin\th_1 \, d\th_1 \wedge d\th_2 & \textrm{for} \quad \cN_2 = S^2
    \\
   d\th_1 \wedge d\th_2 & \textrm{for} \quad \cN_2 = T^2
    \\
   \frac{1}{\ell_2} \, \sinh\th_1 \, d\th_1 \wedge d\th_2 & \textrm{for} \quad \cN_2 = H_2\ .
  \end{cases}
  \, 
\end{equation}
The parameters $c_1 , c_2 , \ell_1 , \ell_2$ are constants which are fixed by 
solving the type-IIB equations of motion. Below we present three such solutions.

\paragraph{Solution I:}
This solution has $\cN_2 = S^2$.  The various constants are
\begin{equation}
 \label{solIAdS6}
 \begin{split}
&  c_2 = c_3 = 0  \, , \quad c_1 = s_1 \sqrt{\frac{\n - 4 \m}{3}} \, , \quad c_4 = s_4 \sqrt{\frac{\n + 8 \m}{3}} \, , 
\\ 
 & \ell_1 = \frac{\n + 2 \m}{6} \, , \quad \ell_2 = \m \, ,
\end{split}
\end{equation}
with $s_{1 , 4} = \pm 1$. The solution is well defined when $c^2_1 , c^2_2 \geqslant 0$ and $\ell_1 , \ell_2 > 0$. This means that $\l$ is bounded as
\begin{equation}
 0 < \l \leqslant 2 - \sqrt{3} \, .
\end{equation}

\paragraph{Solution II:}
The second solution has $\cN_2 = H_2$ and the various constants are
\begin{equation}
\begin{split}
 \label{solIIAdS6}
& c_1 = c_4 = 0 \, , \qquad c_2 = s_2 \sqrt{\frac{\n + 4 \m}{3}}  \, , \qquad c_3 = s_3 \sqrt{\frac{\n - 8 \m}{3}} \, ,
\\
 &  \ell_1 = \frac{\n - 2 \m}{6} \, , \qquad \ell_2 = \m \, ,
\end{split}
\end{equation}
with $s_{2 , 3} = \pm 1$. Requiring that $c^2_2 , c^2_3 \geqslant 0$ and $\ell_1 , \ell_2 > 0$ we find that $\l$ must be restricted as
\begin{equation}
 0 < \l \leqslant 4 - \sqrt{15} \, .
\end{equation}

\paragraph{Solution III:}
When $\l = 0$ one can get solutions with $\cN_2 = T^2$. One of them is the following
\begin{equation}
 \label{solIIIAdS6}
 c_2 = c_3 = 0  \, , \qquad c_1 = s_1 \sqrt{\frac{4}{3 k}} \, , \qquad c_4 = s_4 \sqrt{\frac{4}{3 k}} \, , \qquad \ell_1 = \frac{2}{3 k} \, ,
\end{equation}
with $s_{1 , 4} = \pm 1$. This can also be obtained by taking the $\l = 0$ limit in the solution I and replacing the two-sphere by a torus.

\no
Note that, there is an equivalent solution arising from interchanging the coordinates $y_1$ and $y_2$.


\end{document}